\documentclass[preprint,pre,aps]{revtex4}
\usepackage{graphics}
\usepackage{amsmath,amssymb}
\usepackage[dvips]{graphicx}
\begin{document}
\title{Vapour-liquid phase diagram for an ionic fluid
in a random porous medium
}
 \author{M.F. Holovko, O. Patsahan and T. Patsahan}
\affiliation{Institute for Condensed Matter Physics of the National
Academy of Sciences of Ukraine, 1 Svientsitskii St., 79011 Lviv,
Ukraine}

\date{\today}
\begin{abstract}
We study the vapour-liquid  phase behaviour of an ionic fluid confined in a random porous matrix formed by uncharged hard sphere 
particles.  The ionic fluid is modelled as an equimolar binary mixture of oppositely charged equisized hard spheres, the so-called 
restricted primitive model (RPM).
Considering the matrix-fluid system as a partly-quenched model, we develop a theoretical approach which combines the method of 
collective
variables with the extension of the scaled-particle theory (SPT) for a hard-sphere fluid confined in a disordered hard-sphere matrix.
The approach allows us to formulate the perturbation theory  using the  SPT  for the description of the thermodynamics of the reference
system.
The phase diagrams of the RPM in  matrices  of different porosities  and for different size ratios of matrix and fluid particles
are calculated in the random-phase approximation and also when  the effects of  higher-order correlations between ions are taken into account. 
Both approximations correctly reproduce the basic effects of porous media on the vapour-liquid phase diagram, i.e.,
with a decrease  of porosity the critical point shifts toward lower fluid densities and lower temperatures and the coexistence region
is getting narrower.
For the fixed matrix porosity, both the critical temperature and the critical density
increase with an increase of size of matrix particles and tend to the critical  values of the  bulk RPM.
\end{abstract}
%\pacs{05.70.Fh, 64.60.De, 64.60.F-}

\maketitle

\section{Introduction}

This paper is devoted to the  memory of George Stell who was  one  of the leaders  in the statistical-mechanical theory of the
liquid state for the last fifty years.

The study of the effects of an adsorbent
on phase behaviour of an ionic fluid is far from being solved in spite of its great practical
importance. To our knowledge,  there is no theory capable of  correctly describing the  phase
behaviour of confined ionic liquids even within the simplest models.
By contrast, the  phase transitions and
criticality in  bulk Coulomb-dominated systems  have been  intensively studied for the past decades. The issue was among the  major
 research interests of  George Stell \cite{stell1}.
For reviews of experimental
and theoretical situation see  \cite{Hynnien-Panagiotopoulos,Patsahan_Mryglod_review,Schroer:12} and the references cited therein.
A simple description of systems dominated by Coulomb interactions  is provided by the  restricted primitive model (RPM) that consists
of an equal number  of equisized  positively and negatively charged hard spheres  immersed  into structureless dielectric continuum.
The RPM undergoes a
vapour-liquid-like phase transition at low temperatures and low densities \cite{stell1,levin-fisher,Cai-Mol1,patsahan_ion}.

The existing theoretical approaches consider fluids in disordered confinements as a partly-quenched system
in which some of the degrees of freedom are quenched while the others are annealed \cite {Madden88}.
In this case, statistical-mechanical averages used for calculations of
thermodynamic properties become double ensemble averages: the first average is taken over all degrees of freedom of annealed
particles  keeping the quenched  particles fixed,
and the other average is performed  over all realizations of a matrix. The application of the replica method made
it possible to relate
the matrix averaged quantities to the thermodynamic quantities of a corresponding fully equilibrated model, called
the replicated model \cite{Given:92,Given_Stell:92,Given95,Rosinberg:94}.
The major part of theoretical studies
of partly-quenched systems
containing charges was mainly focused on structural  and  thermodynamic properties within the framework
of the replica Ornstein-Zernike (ROZ) theory (see  review~\cite{Hribar_Lee:11} and the references cited therein). The studies of the
phase behaviour of partly-quenched models using the ROZ were concerned with the systems characterized by short-range interactions
\cite{Trokhymchuk:95,Kahl:00,Kahl:01}.
However, unlike  bulk fluids,
no analytical result has been obtained  from the ROZ integral equation approach even for   a hard-sphere
fluid  confined  in a hard-sphere matrix, being the  model of particular importance for the development of perturbation theories.

Recently, a pure analytical approach has been developed to describe the thermodynamics of liquids confined
in disordered porous materials composed of rigid particles
\cite{Holovko_Patsahan_Dong:12,Holovko_Patsahan_Dong:13,Holovko_Shmotolokha_Patsahan:14,Holovko_Shmotolokha_Patsahan_chapter}.
This approach is based on the  scaled-particle theory (SPT) \cite{Reiss:59} and is capable of rather correctly  reproducing thermodynamic
properties of confined fluids. Although the SPT is
limited to hard-core interaction between molecules, it can be used for the description of  a reference
system within the perturbation theory to study different complex systems with attractive, repulsive, and associative interactions
\cite{Kalyuzhnyi_Holovko_Patsahan_Cummings:14,Holovko_Patsahan_Shmotolokha:15}.

The purpose of the present  paper is to   study the vapour-liquid phase behaviour of the RPM confined
in a random porous matrix
formed by uncharged   particles. We start with a  more general model when  the interaction potentials  between the two matrix particles
and  between the ion and the matrix
particle  include a short-range attraction/repulsion in
addition to  a hard-core repulsion. Furthermore, the ions and the matrix particles  differ in size.
Following the formalism originally proposed by Madden and Gland  for a one-component system in a disordered porous
matrix  \cite{Madden88},  we
consider  our matrix-fluid system as a partly-quenched model. Combining the collective variable (CV) method 
\cite{Yukhnovskii_Holovko,Cai-Pat-Mryg:05,Cai-Pat-Mryg,Patsahan_Mryglod_Caillol} with the SPT, we develop a 
theoretical approach  which  allows us to formulate the perturbation 
theory and to  treat the model of an uncharged  hard-sphere fluid
in an uncharged  hard-sphere matrix as a reference system. It should be noted that the CV method is a  useful tool 
for the study of  phase transitions  in  systems 
with Coulomb interactions \cite{patsahan_ion,patsahan-mryglod-patsahan:06,Patsahan_Patsahan:10}.
Using the replica trick, we derive an   expression for the
 grand partition function (GPF) of a wholly equilibrated ($2s+1$)-component system in the form of a functional integral.
In the limit $s\rightarrow 0$,  
we obtain  explicit expressions
for the Helmholtz free energy
and the corresponding chemical potential of an ionic fluid in
the random-phase approximation (RPA).   For the  particular
case  of neglecting matrix-matrix and matrix-fluid interactions outside the hard core,  we  calculate  vapour-liquid coexistence 
curves of the RPM confined in hard-sphere matrices
of different porosity.
It should be noted that  for the unconfined (bulk) state, the RPA, just as the standard  Debye-H\"{u}ckel theory and
the mean-spherical approximation \cite{stell1,levin-fisher},  predicts the  critical point parameters
 widely deviating from the  simulation data \cite{Hynnien-Panagiotopoulos}.
In order to correct the situation we  derive an explicit equation for the vapour-liquid
spinodal  using the method proposed in \cite{patsahan_ion} for the bulk RPM.  The method produces
vapour-liquid critical parameters of the RPM in particular, the critical temperature,
which are in good agreement  with simulation results. The resulting  equation for the
spinodal includes   isothermal compressibility  of the reference system and its first and second derivatives. In this approximation,
the trends of the critical parameters
(maximum points of the spinodals) with the matrix characteristics are the same as those obtained in the RPA.

The paper is arranged as follows. In section~2 we present a theoretical  formalism.
In section~3 we calculate  the vapour-liquid phase diagrams of the RPM confined in a disordered hard-sphere matrix  in two
approximations, i.e., in the RPA  and in the approximation
which takes into account the effects of the higher-order correlations.  The discussion  of the effects of  matrix properties on
the phase diagrams  is also presented  here. We draw conclusions in section 4.

 \section{Theoretical background}

\subsection{Model}
Our matrix/ionic-fluid system contains two subsystems: the first one (i.e., matrix) is composed of particles quenched or frozen in place,
while
the second subsystem is  an annealed  (or allowed to equilibrate) binary ionic fluid which is in equilibrium with the matrix.
It is assumed that
the matrix particles  were quenched into an equilibrium configuration corresponding to the Gibbs distribution associated with  a
pairwise  interaction potential $u_{00}(x)$.
The ionic fluid is treated as the RPM.
The interactions
in the matrix/ionic-fluid system  are described by  a set of the  pairwise interaction potentials:
$u_{00}(x)$, $u_{++}(x)=u_{--}(x)$, $u_{+-}(x)=u_{-+}(x)$,  $u_{0+}(x)=u_{+0}(x)$, $u_{0-}(x)=u_{-0}(x)$,
where the subscript $0$ refers to the matrix particles and the subscripts $+,-$ refer to the ions. We also assume that
$u_{0+}(x)=u_{0-}(x)$.

The grand canonical potential of the system is
given by \cite{Madden88}
\begin{equation*}
-\beta\overline{\Omega}_{1}=\frac{1}{\Xi_{0}}\sum_{N_{0}\geq 0}\frac{z_{0}^{N_{0}}}{N_{0}!}
\int {\rm d}{\mathbf q}^{N_{0}}
\exp[-\beta H_{00}({\mathbf q}^{N_{0}})]\ln\Xi_{1}({\mathbf q}^{N_{0}}),
%\label{2.1}
\end{equation*}
where
\begin{equation*}
\Xi_{0}=\sum_{N_{0}\geq 0}\frac{z_{0}^{N_{0}}}{N_{0}!}\int {\rm d}{\mathbf q}^{N_{0}}
\exp[-\beta H_{00}({\mathbf q}^{N_{0}})]
%\label{2.1a}
\end{equation*}
is the GPF  for the prequench medium,
${\mathbf q}^{N_{0}}={{\mathbf q}_{1},{\mathbf q}_{2},\ldots,{\mathbf q}_{N_{0}}}$ denotes the positions of the matrix particles,
$z_{0}$ is the activity of the matrix particles prior  to quenching,   and
$H_{00}({\mathbf q}^{N_{0}})$ is the potential energy  of the matrix particle configuration.
The matrix-dependent  GPF $\Xi_{1}$ has the form:
 \begin{eqnarray*}
\Xi_{1}({\mathbf q}^{N_{0}})=\sum_{N_{+}\ge 0}\sum_{N_{-}\ge 0}\frac{z_{+}^{N_{+}}}{N_{+}!}\frac{z_{-}^{N_{-}}}{N_{-}!}
\int {\rm d}{\mathbf r}^{N_{+}}{\rm d}{\mathbf r}^{N_{-}}\exp\{-\beta[H_{0+}({\mathbf r}^{N_{+}},{\mathbf q}^{N_{0}})
\nonumber \\
+H_{0-}({\mathbf r}^{N_{-}},{\mathbf q}^{N_{0}})
+H_{++}({\mathbf r}^{N_{+}})+H_{--}({\mathbf r}^{N_{-}})+H_{+-}({\mathbf r}^{N_{+}},{\mathbf r}^{N_{-}})]\},
%\label{Xi1}
\end{eqnarray*}
where  $z_{A}$ is the activity of the ions of  species  $A$ ($A=+,-$),
${\mathbf r}^{N_{A}}={{\mathbf r}_{1}^{A},{\mathbf r}_{2}^{A},\ldots,{\mathbf r}_{N_{A}}^{A}}$ denotes the positions of the ions,
$H_{ij}$ describe the potential energy of  the $N_{1}=N_{+}+N_{-}$ ($N_{+}=N_{-}$) ions
in the presence of  matrix obstacles, $\beta=1/k_{\rm{B}}T$ is the reciprocal temperature.

Following \cite{Given95}, we make use of the replica trick, which consists in replacing the logarithm with an exponential. As a result, we have
\begin{equation}
-\beta\overline{\Omega}_{1}=\beta V\overline{P}=\lim_{s \to 0}\frac{d}{ds}\ln \Xi^{\rm{\rm{rep}}}(s),
\label{2.2}
\end{equation}
where
\begin{eqnarray}
&&\Xi^{\rm{rep}}(s)=\sum_{N_{0}\ge 0}\sum_{N_{1}^{+}\ge 0}\ldots
\sum_{N_{s}^{+}\ge 0}\sum_{N_{1}^{-}\ge 0}\ldots
\sum_{N_{s}^{-}\ge 0}\frac{z_{0}^{N_{0}}}{N_{0}!}\prod_{\alpha=1}^{s}\frac{z_{+,\alpha}^{N_{\alpha}^{+}}
z_{-,\alpha}^{N_{\alpha}^{-}}}{N_{\alpha}^{+}!N_{\alpha}^{-}!}
\nonumber \\
\times
&&\int {\rm d}{\mathbf q}^{N_{0}} \prod_{\alpha=1}^{s}
{\rm d}{\mathbf r}_{\alpha}^{N_{\alpha}^{+}}
{\rm d}{\mathbf r}_{\alpha}^{N_{\alpha}^{-}}
\exp\{-\beta H_{00}-\beta\sum_{\alpha=1}^{s}[H_{0+}^{\alpha}+H_{0-}^{\alpha}]
\nonumber \\
&&-\beta\sum_{\alpha,\beta=1}^{s}[H_{++}^{\alpha\beta}
+H_{+-}^{\alpha\beta}
+H_{--}^{\alpha\beta}] \}.
\label{Ksi_rep}
\end{eqnarray}
$\Xi^{\rm{\rm{rep}}}(s)$ is the GPF of a fully equilibrated  ($2s+1$)-component mixture, consisting
of the matrix  and of $s$ identical copies or replicas  of the two-component ionic fluid.   Each pair of particles has the same
pairwise interaction in this replicated system as in the partly quenched  model except that a pair of ions from different
replicas has no interaction
\begin{equation*}
 H_{00}=\sum_{i<j}^{N_{0}}u_{00}(|{\mathbf q}_{i}-{\mathbf q}_{j}|), \qquad
 H_{0A}^{\alpha}=\sum_{i<j}^{N_{0},N_{\alpha}^{A}}u_{0A}^{\alpha}({\mathbf r}_{\alpha,i}^{A}-{\mathbf q}_{j}|),
%\label{H00}
\end{equation*}
\begin{equation*}
 H_{AB}^{\alpha\beta}=\sum_{i<j}^{N_{\alpha}^{A},N_{\beta}^{B}}u_{AB}^{\alpha\beta}(|{\mathbf r}_{\alpha,i}^{A}-{\mathbf r}_{\beta,j}^{B}|)
 \delta_{\alpha\beta}.
 %\label{H0A-AB}
\end{equation*}
In the above equations,   Greek indices $\alpha,\beta$  denote replicas and Latin indices $A,B$ denote ionic species,
${\mathbf r}_{\alpha}^{N_{\alpha}^{A}}=
{{\mathbf r}_{\alpha,1}^{A},{\mathbf r}_{\alpha,2}^{A},\ldots,{\mathbf r}_{\alpha,N_{A}}^{A}}$ is used for the set of charged
particles of species $A$ in  replica $\alpha$.
Since the replicated system is an equilibrium system, one can get the Helmholtz free energy  using the Legendre transform
\begin{equation}
 F^{\rm{rep}}(s)=-\ln\Xi^{\rm{rep}}(s)+\mu_{0}^{\rm{rep}} N_{0}^{\rm{rep}}+\sum_{\alpha=1}^{2s}\mu_{\alpha}^{\rm{rep}} N_{\alpha}^{\rm{rep}}.
\label{A_rep}
\end{equation}
Then,  the free energy $\overline{F}$ of the partly quenched system can be found from the equation
\begin{equation}
-\beta\overline{F}=\lim_{s \to 0}\frac{d}{ds}(-\beta F^{\rm{rep}}).
\label{F_slim}
\end{equation}

\subsection{The method of collective variables}

We assume that all the pairwise potentials  between the different particles can be split into a reference part (index ``r'') and
a perturbation part (index ``p'')
\begin{eqnarray}
 u_{ij}(x)=u_{ij}^{(r)}(x)+u_{ij}^{(p)}(x),
 \label{split}
 \end{eqnarray}
where $u_{ij}^{(r)}(x)$ is a potential of a short-range repulsion which, generally, describes the mutual impenetrability of the particles,
while  $u_{ij}^{(p)}(x)$ mainly describes the behaviour  at moderate and large distances.
The system in which the particles interact via the potentials $u_{ij}^{(r)}(x)$ is regarded as the reference system, $u_{ij}^{(r)}(x)$
is specified in the form of the hard-sphere potential. We assume
that the thermodynamic and structural properties of the reference system are known.
The interactions connected with  potentials $u_{ij}^{(p)}(x)$  are considered
in the phase space of CVs.

Using the CV method we can get the exact functional representation of the GPF
(\ref{Ksi_rep})  \cite{Cai-Pat-Mryg,Patsahan_Mryglod_Caillol}
\begin{eqnarray}
\Xi^{\rm{rep}}(s)=\int ({\rm d}\rho_{0})({\rm d}\omega_{0})({\rm d}\rho_{A}^{\alpha})({\rm d}\omega_{A}^{\alpha})
 \exp\left[-\frac{\beta}{2V}\sum_{{\mathbf k}}
 \tilde{u}_{00}^{(p)}(k)\rho_{{\mathbf k},0}\rho_{-{\mathbf k},0}\right. \nonumber\\
\left.
-\frac{\beta}{V}\sum_{\alpha}\sum_{A}
\sum_{{\mathbf k}}\tilde{u}_{0A}^{\alpha(p)}(k)\rho_{{\mathbf k},0}\rho_{-{\mathbf k},A}^{\alpha}
-\frac{\beta}{2V}\sum_{\alpha}\sum_{A,B}\sum_{{\mathbf k}}\tilde{u}_{AB}^{\alpha\alpha(p)}(k)\rho_{{\mathbf k},A}^{\alpha}
\rho_{-{\mathbf k},B}^{\alpha}\right.
\nonumber\\
 \left.
+{\rm i}\sum_{{\mathbf k}}\omega_{{\mathbf k},0}\rho_{{\mathbf k},0}+{\rm i}\sum_{\alpha}\sum_{A}\sum_{{\mathbf k}}
\omega_{{\mathbf k},A}^{\alpha}\rho_{{\mathbf k},A}^{\alpha}+\ln\Xi^{r}[\bar\nu_{0}-{\rm i}\omega_{0}, \bar\nu_{A}^{\alpha}
-{\rm i}\omega_{A}^{\alpha}]\right].
\label{Xi_rep}
\end{eqnarray}
Here, the following notations are introduced: $\rho_{{\mathbf k},0}=\rho_{{\mathbf k},0}^{(c)}-{\rm i}\rho_{{\mathbf k},0}^{(s)}$
and $\rho_{{\mathbf k},A}^{\alpha}=\rho_{{\mathbf k},A}^{\alpha (c)}-{\rm i}\rho_{{\mathbf k},A}^{\alpha(s)}$ are the
CVs which describe the  fluctuation modes
of the  number  of the matrix and fluid species, respectively,  indices $c$ and $s$ denote real and imaginary parts of
CVs,  $\omega_{{\mathbf k},0}$ and $\omega_{{\mathbf k},A}^{\alpha}$ are
conjugate to CVs $\rho_{{\mathbf k},0}$ and $\rho_{{\mathbf k},A}^{\alpha}$, respectively. $({\rm
d}\rho_{0})$ and $({\rm d}\rho_{A}^{\alpha})$ denote  volume elements of the CV
phase space
\begin{displaymath}
({\rm d}\rho_{0})={\rm d}\rho_{0,0}{\prod_{\mathbf
k\not=0}}' {\rm d}\rho_{\mathbf k,0}^{(c)}{\rm d}\rho_{\mathbf
k,0}^{(s)}, \quad ({\rm d}\rho_{A}^{\alpha})=\prod_{\alpha}\prod_{A=+,-}{\rm
d}\rho_{0,A}^{\alpha}{\prod_{\mathbf k\not=0}}' {\rm d}\rho_{\mathbf
k,A}^{\alpha (c)}{\rm d}\rho_{\mathbf k,A}^{\alpha (s)},
\end{displaymath}
the product over ${\mathbf k}$ is performed in the upper
semi-space ($
\rho_{-\mathbf k,0}=\rho_{\mathbf k,0}^{*}$, $\rho_{-\mathbf k,A}^{\alpha}=\rho_{\mathbf k,A}^{\alpha *}$).
The same kind of relationships  hold for $({\rm d}\omega_{0})$ and $({\rm d}\omega_{A}^{\alpha})$.
Coefficients
$\tilde{u}_{00}^{(p)}(k)$, $\tilde{u}_{0A}^{\alpha(p)}(k)$ and $\tilde{u}_{AB}^{\alpha\alpha(p)}(k)$ are the
Fourier transforms of the corresponding interaction potentials.

$\Xi^{r}[\bar\nu_{0}-{\rm i}\omega_{0}, \bar\nu_{A}^{\alpha}
-{\rm i}\omega_{A}^{\alpha}]$ is the GPF
of the reference system  with the renormalized chemical potentials
\begin{eqnarray}
\bar\nu_{0}=\nu_{0}+\frac{\beta}{2V}\sum_{{\mathbf
k}}\tilde{u}_{00}^{(p)}(k), \qquad
\bar\nu_{A}^{\alpha}=\nu_{A}^{\alpha}+\frac{\beta}{2V}\sum_{{\mathbf
k}}\tilde u_{AA}^{\alpha\alpha(p)}(k),
\label{bar_nu}
\end{eqnarray}
where $\nu_{0}=\beta\mu_{0}-\ln\Lambda^{3}$ and $\nu_{A}^{\alpha}=\beta\mu_{A}^{\alpha}-\ln\Lambda^{3}$
are the dimensionless chemical potentials of the corresponding species,
$\Lambda$ is
the  de Broglie thermal wavelength.
In order to formulate the perturbation theory we apply  a cumulant theorem  to $\ln\Xi^{r}$ \cite{Patsahan_Mryglod_Caillol}.
As a result, equation~(\ref{Xi_rep}) reads
\begin{eqnarray}
\Xi^{\rm{rep}}(s)=\Xi^{\rm{mf}}[\tilde{\nu}_{0},\tilde{\nu}_{A}^{\alpha}]\int ({\rm d}\rho_{0})({\rm d}\omega_{0})
({\rm d}\rho_{A}^{\alpha})({\rm d}\omega_{A}^{\alpha})
\exp\left[-\frac{\beta}{2}\sum_{{\mathbf k}}\widehat{U}(k)\widehat{\rho}_{{\mathbf k}}\widehat{\rho}_{-{\mathbf k}}\right.
\nonumber\\
\left.
+{\rm i}\sum_{{\mathbf k}}\widehat{\omega}_{{\mathbf k}}\widehat{\rho}_{{\mathbf k}}
+\sum_{n\geq 2}\frac{(-{\rm i})^{n}}{n!}\sum_{{\mathbf{k}}_{1},\ldots,{\mathbf{k}}_{n}}\widehat{{\mathfrak{M}}}_{n}\widehat{\omega}_{\mathbf k_{1}}\widehat{\omega}_{\mathbf k_{2}}\ldots\widehat{\omega}_{\mathbf k_{n}}\delta_{{\bf{k}}_{1}+\ldots +{\bf{k}}_{n}}
\right].
\label{Xi_matrix}
\end{eqnarray}
where $\Xi^{\rm{mf}}$ is the mean-field part of the GPF
\begin{equation*}
\Xi^{\rm{mf}}=\Xi^{r}[\tilde{\nu}_{0},\tilde{\nu}_{A}^{\alpha}]\exp\left\lbrace\langle N_{0}\rangle_{r}\left[\frac{\beta}{2}\bar\rho_{0}\tilde{u}_{00}^{(p)}(0)
+\sum_{\alpha}\sum_{A}\beta\bar\rho_{A}^{\alpha}\tilde{u}_{0A}^{\alpha(p)}(0) \right]\right\rbrace,
%\label{Xi_mf}
\end{equation*}
and $\Xi^{r}$  depends on
\begin{eqnarray}
\tilde{\nu}_{0}=\bar\nu_{0}-\bar\rho_{0}\beta\tilde{u}_{00}^{(p)}(0)-\sum_{\alpha}\sum_{A}\bar\rho_{A}^{\alpha}
\beta\tilde{u}_{0A}^{\alpha(p)}(0),
\nonumber \\
\tilde{\nu}_{A}^{\alpha}=\bar\nu_{A}^{\alpha}-\bar\rho_{0}\beta\tilde{u}_{0A}^{\alpha(p)}(0)-
\sum_{B}\bar\rho_{B}^{\alpha}\beta\tilde{u}_{AB}^
{\alpha\alpha(p)}(0),
\label{tilde_nu}
\end{eqnarray}
$\bar\rho_{0}=\langle N_{0}\rangle_{r}/V$, $\bar\rho_{A}^{\alpha}=\langle N_{A}^{\alpha}\rangle_{r}/V$, $\langle\ldots\rangle_{r}$
indicates the  average taken over the reference system.

$\widehat{U}(k)$ denotes a symmetric $(2s+1)\times(2s+1)$
 matrix of elements:
\begin{eqnarray}
u_{11}&=&\widetilde{u}_{00}^{(p)}(k)=\widetilde{\varphi}_{00}(k) \nonumber \\
u_{1i}&=&u_{i1} =\widetilde{u}_{0A}^{\alpha(p)}(k)=\widetilde{\varphi}_{0I}(k), \qquad
(i\geq 2),\nonumber \\
u_{ii}&=&\widetilde{u}_{AA}^{\alpha\alpha(p)}(k)=\widetilde{\varphi}^{C}(k), \qquad
(i\geq 2), \nonumber \\
u_{ij}&=&-\widetilde{u}_{AA}^{\alpha\alpha(p)}(k)=-\widetilde{\varphi}^{C}(k), \qquad
(2\leq i\leq 2s,\, j=i+s), \nonumber \\
u_{ij}&=&0, \qquad (2\leq i\leq 2s,\, j\neq i+s),
\label{matrix_poten}
\end{eqnarray}
$\widetilde{\varphi}^{C}(k)$ is the Fourier transform of the Coulomb potential $\varphi^{C}(r)=q^{2}/(\epsilon r)$.
$\widehat{\rho}_{{\mathbf k}}$  indicates a column vector of elements $\rho_{{\mathbf k},0}$, $\rho_{{\mathbf k},+}^{1}$,
$\ldots$, $\rho_{{\mathbf k},+}^{s}$, $\rho_{{\mathbf k},-}^{1}$,
$\ldots$, $\rho_{{\mathbf k},-}^{s}$ and $\widehat{\omega}_{\mathbf k}$ is a row vector of elements $\omega_{{\mathbf k},0}$,
$\omega_{{\mathbf k},+}^{1}$,
$\ldots$, $\omega_{{\mathbf k},+}^{s}$, $\omega_{{\mathbf k},-}^{1}$,
$\ldots$, $\omega_{{\mathbf k},-}^{s}$.
$\widehat{{\mathfrak{M}}}_{n}$ is a  symmetric
$\underbrace{(2s+1)\times(2s+1)\times\ldots\times(2s+1)}_{n}$  matrix   whose elements are cumulants: the $n$th cumulant
coincides with the Fourier transform of the $n$-particle  connected correlation function of the reference
system.

\subsection{Free energy of the ionic model in a disordered matrix: the random phase approximation}

We consider the Gaussian approximation of the GPF putting $n=2$ in the cumulant expansion on the right-hand side
of (\ref{Xi_matrix}).
In this case, $\widehat{{\mathfrak{M}}}_{2}$ is a symmetric $(2s+1)\times(2s+1)$ matrix. In the reference system,
considered in this paper,
the interaction potentials between the two fluid particles are equal (hard-sphere diameters of the ions are equal,
$\sigma_{+}=\sigma_{-}=\sigma_{1}$) but, in general, they  differ from those between
the two matrix particles ($\sigma_{1}\neq\sigma_{0
}$).  In this case, the elements of matrix    $\widehat{{\mathfrak{M}}}_{2}$ are as follows:
\begin{eqnarray*}
m_{11}&=&{\mathfrak{M}}_{00}(k), \quad
 m_{1i}=m_{i1}= {\mathfrak{M}}_{01}(k),  \quad (i\geq 2) \nonumber \\
 m_{ii}&=&{\mathfrak{M}}_{11}(k), \quad (i\geq 2) \nonumber \\
 m_{ij}&=&{\mathfrak{M}}_{12}(k), \quad (i,j\geq 2, \, i\neq j),
\end{eqnarray*}
where
${\mathfrak{M}}_{ij}(k)=\langle N_{i}\rangle_{r}[\delta_{ij}+\bar\rho_{j}\widetilde{h}_{ij}^{r}(k)]$,
the subscript $0$ refers to the matrix particles  and the  subscripts $1$ and $2$ refer to the fluid,
$\widetilde{h}_{ij}^{r}(k)$ is the Fourier transform of the total correlation function in the reference system.
The determinant of  matrix $\widehat{{\mathfrak{M}}}_{2}$ has the form:
\begin{equation}
\det[\widehat{{\mathfrak{M}}}_{2}(s)]=({\mathfrak{M}}_{11}-{\mathfrak{M}}_{12})^{2s-1}\Delta_{1},
\label{det_m2}
\end{equation}
with
$\Delta_{1}=2s({\mathfrak{M}}_{12}{\mathfrak{M}}_{00}
-{\mathfrak{M}}_{01}^{2})+{\mathfrak{M}}_{00}({\mathfrak{M}}_{11}-{\mathfrak{M}}_{12})$. For $\sigma_{+}=\sigma_{-}$, we have
${\mathfrak{M}}_{11}-{\mathfrak{M}}_{12}=\langle N_{1}\rangle_{r}/2$.

After integration in (\ref{Xi_matrix}), we arrive at the GPF of the replicated system in the Gaussian
approximation
\begin{eqnarray*}
 \frac{1}{V}\ln\Xi_{\rm{G}}^{\rm{rep}}(s)&=&\frac{1}{V}\ln\Xi^{r}+\frac{\beta}{2}(\bar\rho_{0})^{2}\widetilde{\varphi}_{00}(0)
+s\beta\bar\rho_{0}\bar\rho_{1}\widetilde{\varphi}_{0I}(0)
\nonumber \\
&&-\frac{1}{2V}\sum_{{\mathbf{k}}}\ln\left[\det(\widehat{{\mathfrak{M}}}_{2}\widehat{C}_{2})\right],
%\label{Xi_G}
\end{eqnarray*}
where $\bar\rho_{1}=\bar\rho_{+}+\bar\rho_{-}$ and
$\widehat{C}_{2}$ denotes a symmetric $(2s+1)\times(2s+1)$  matrix of the Fourier transforms of the two-particle
direct correlation functions in the
Gaussian approximation. Its elements  are as follows:
\begin{eqnarray}
 c_{11}&=&\widetilde{C}_{00}(s)=\frac{\beta}{V}\widetilde{\varphi}^{00}(k)+\frac{{\mathfrak{M}}_{11}+(2s-1){\mathfrak{M}}_{12}}{\Delta_{1}}, \nonumber \\
 c_{1i}&=&c_{i1}=\widetilde{C}_{01}(s)=\frac{\beta}{V}\widetilde{\varphi}^{0I}(k)-\frac{{\mathfrak{M}}_{01}}{\Delta_{1}}, \quad (i\geq 2),\nonumber \\
 c_{ii}&=&\widetilde{C}_{11}(s)=\frac{\beta}{V}\widetilde{\varphi}^{C}(k)+\frac{{\mathfrak{M}}_{00}}{\Delta_{1}}
+\frac{(2s-1)({\mathfrak{M}}_{00}{\mathfrak{M}}_{12}
 -{\mathfrak{M}}_{01}^{2})}{\Delta_{2}}, \quad (i\geq 2),\nonumber \\
 c_{ij}&=&\widetilde{C}_{12}(s) =-\frac{\beta}{V}\widetilde{\varphi}^{C}(k)-\frac{{\mathfrak{M}}_{00}{\mathfrak{M}}_{12}
 -{\mathfrak{M}}_{01}^{2}}{\Delta_{2}}, \quad   (2\leq i\leq 2s, \,\,j=i+s),\nonumber \\
 c_{ij}&=&\widetilde{C}_{12}^{\alpha\beta}(s)=-\frac{{\mathfrak{M}}_{00}{\mathfrak{M}}_{12}
 -{\mathfrak{M}}_{01}^{2}}{\Delta_{2}}, \quad   (2\leq i\leq 2s, \,\,j\neq i+s),
 \label{cij}
 \end{eqnarray}
where
$\Delta_{2}=({\mathfrak{M}}_{11}-{\mathfrak{M}}_{12})\Delta_{1}$.  Then, the determinant of matrix $\widehat{C}_{2}$ has the form:
\begin{eqnarray}
\det[\widehat{C}_{2}(s)]=\left(\widetilde{C}_{11}-\widetilde{C}_{12}\right)^{s}\left(\widetilde{C}_{11}+
\widetilde{C}_{12}-2\widetilde{C}_{12}^{\alpha\beta}\right)^{s-1}
\nonumber \\
\times\left\lbrace \widetilde{C}_{00}\left[
\widetilde{C}_{11}+
\widetilde{C}_{12}+2(s-1)\widetilde{C}_{12}^{\alpha\beta}\right]-2s(\widetilde{C}_{01})^{2}\right\rbrace.
\label{det_C2}
\end{eqnarray}

Taking into account (\ref{det_m2}) and (\ref{cij})-(\ref{det_C2}), we get the following expression for  $\ln\Xi_{\rm{G}}^{\rm{rep}}$
\begin{eqnarray}
&&\frac{1}{V}\ln\Xi_{\rm{G}}^{\rm{rep}}(s)=\frac{1}{V}\ln\Xi^{r}+\frac{(\bar\rho_{0})^{2}}{2}\beta\widetilde{\varphi}_{00}(0)
+s\bar\rho_{0}\bar\rho_{1}\beta\widetilde{\varphi}_{0I}(0)
\nonumber \\
&&-\frac{s}{2V}\sum_{{\mathbf{k}}}\ln\left[
 1+\bar\rho_{1}\beta\widetilde{\varphi}^{C}(k)\right]-
 \frac{1}{2V}\sum_{{\mathbf{k}}}\ln\left[1+\bar\rho_{0}\beta\widetilde{\varphi}^{00}(k)S_{00}\right. \nonumber \\
&&\left.
  +
 4s\sqrt{\bar\rho_{0}\bar\rho_{1}}\beta\widetilde{\varphi}^{0I}(k)S_{01}
 -2s\bar\rho_{0}\bar\rho_{1}
(\beta\widetilde{\varphi}^{0I}(k))^{2}\overline{\Delta}_{1}\right],
\label{Omega_rep}
\end{eqnarray}
where $S_{ij}={\mathfrak{M}}_{ij}/\sqrt{\langle N_{i}\rangle_{r}\langle N_{j}\rangle_{r}}$ and
$\overline{\Delta}_{1}=\Delta_{1}/\langle N_{0}\rangle_{r}\langle N_{1}\rangle_{r}$.

Using (\ref{A_rep}), (\ref{bar_nu}) and (\ref{tilde_nu}) we arrive at the Helmholtz free energy per volume
of the replicated system, $\beta f=\beta F^{\rm{rep}}/V$,
in the RPA
\begin{eqnarray*}
&&\beta f_{\rm{RPA}}=\beta f^{r}(s)
 -\frac{\rho_{0}^{\rm{rep}}}{2V}\sum_{{\mathbf{k}}}\beta\widetilde{\varphi}^{00}(k)
  -s\frac{\rho_{1}^{\rm{rep}}}{2V}\sum_{{\mathbf{k}}}\beta\widetilde{\varphi}^{C}(k) \nonumber \\
 && +\frac{(\rho_{0}^{\rm{rep}})^{2}}{2}\beta\widetilde{\varphi}^{00}(0)
+ s\rho_{0}^{\rm{rep}}\rho_{1}^{\rm{rep}}\beta\widetilde{\varphi}^{0I}(0)
+\frac{s}{2V}\sum_{{\mathbf{k}}}\ln\left[
 1+\rho_{1}^{\rm{rep}}\beta\widetilde{\varphi}^{C}(k)\right]\nonumber \\
&&+\frac{1}{2V}\sum_{{\mathbf{k}}}\ln\left[1+\rho_{0}^{\rm{rep}}\beta\widetilde{\varphi}^{00}(k)S_{00}
 +4s\sqrt{\rho_{0}^{\rm{rep}}\rho_{1}^{\rm{rep}}}\beta\widetilde{\varphi}^{0I}(k)S_{01}\right. \nonumber \\
&&\left.
 -2s\rho_{0}^{\rm{rep}}\rho_{1}^{\rm{rep}}(\beta\widetilde{\varphi}^{0I}(k))^{2}\overline{\Delta_{1}}\right],
%\label{f_rpa}
\end{eqnarray*}
where $f^{r}$
is the free energy of the reference system.
$\rho_{0}^{\rm{rep}}$ and $\rho_{1}^{\rm{rep}}$  denote the number densities of the matrix  and fluid particles,  respectively.

Finally, taking a replica limit (\ref{F_slim}), we find the following expression for the RPA free energy of a model ionic fluid in a
disordered matrix
\begin{eqnarray}
\beta \overline{f}_{\rm{RPA}}=\beta\overline{f}^{r}
  -\frac{1}{2V}\sum_{{\mathbf{k}}}\beta\widetilde{\varphi}^{00}(k)\left.\frac{d}{ds}\right\vert_{s=0}\rho_{0}^{\rm{rep}}
  -\frac{\rho_{1}}{2V}\sum_{{\mathbf{k}}}\beta\widetilde{\varphi}^{C}(k) \nonumber \\
 +\beta\widetilde{\varphi}^{00}(0) \rho_{0}\left.\frac{d}{ds}\right\vert_{s=0}\rho_{0}^{\rm{rep}}
+ \rho_{0}\rho_{1}\beta\widetilde{\varphi}^{0I}(0)
+\frac{1}{2V}\sum_{{\mathbf{k}}}\ln\left[
 1+\rho_{1}\beta\widetilde{\varphi}^{C}(k)\right]
\nonumber \\
 +\frac{1}{2V}\sum_{{\mathbf{k}}}\left\lbrace\left[\beta\widetilde{\varphi}^{00}(k)\left.\frac{d}{ds}\right\vert_{s=0}(\rho_{0}^{\rm{rep}}S_{00})
 +4\beta\widetilde{\varphi}^{0I}(k)\sqrt{\rho_{0}\rho_{1}}\left.S_{01}\right\vert_{s=0}\right.\right. \nonumber \\
\left.\left.
 -(\beta\widetilde{\varphi}^{0I}(k))^{2}\rho_{0}\rho_{1}\left.S_{00}\right\vert_{s=0}\right]/\left[1+
 \beta\widetilde{\varphi}^{00}(k)\rho_{0}\left.S_{00}\right\vert_{s=0}\right]\right\rbrace.
\label{f_PQ}
 \end{eqnarray}
In the above equation,  the notations  $\rho_{0}=\left.\rho_{0}^{\rm{rep}}\right\vert_{s=0}$ and
$\rho_{1}=\left.\rho_{1}^{\rm{rep}}\right\vert_{s=0}$ are introduced.

Hereafter, we put $\widetilde{\varphi}^{00}(k)=0$ and $\widetilde{\varphi}^{0I}(k)=0$ which corresponds to neglecting the perturbative
parts of the interaction potentials between the matrix particles and between the matrix particles and the ions (see (\ref{split})
and (\ref{matrix_poten})).
As a result,  (\ref{f_PQ}) reduces to the form:
\begin{eqnarray*}
 \beta \overline{f}_{\rm{RPA}}-\beta\overline{f}^{r}=
 -\frac{\rho_{1}}{2V}\sum_{{\mathbf{k}}}\beta\widetilde{\varphi}^{C}(k)+\frac{1}{2V}\sum_{{\mathbf{k}}}\ln\left[
 1+\rho_{1}\beta\widetilde{\varphi}^{C}(k)\right].
%\label{f_PQ_0}
 \end{eqnarray*}
The term  on the right-hand side of the above equation arises from the electrostatic interactions
between the ions. It is of the same form as
the corresponding contribution to the free energy of the  RPM in the bulk \cite{patsahan-mryglod-patsahan:06}.
The main difference between the bulk RPM and the RPM confined in a hard-sphere matrix is given by the term $\overline{f}^{r}$ 
describing the contribution from the reference system.
Accordingly, the perturbative part of the  chemical potential ($\mu_{1}=\mu_{+}=\mu_{-}$) of the ions
 in the presence of a hard-sphere matrix   reads
\begin{eqnarray}
 \overline{\mu}_{1}^{\rm{RPA}}-\overline{\mu}_{1}^{r}=-\frac{1}{2V}\sum_{{\mathbf{k}}}\widetilde{\varphi}^{C}(k)-
 \frac{\beta^{-1}}{2V}\sum_{{\mathbf{k}}}\widetilde{g}(k),
 \label{nu_rpa}
\end{eqnarray}
where
\begin{equation}
\widetilde{g}(k)=-\frac{\beta\widetilde{\varphi}^{C}(k)}{1+\rho_{1}\beta\widetilde{\varphi}^{C}(k)}
\label{g_k}
\end{equation}
and $\overline{\mu}_{1}^{r}$ is a part of the  chemical potential connected with in the reference system.

\subsection{Beyond the random-phase approximation}

Although the RPA  gives
a good qualitative description of the vapour-liquid phase behaviour of the RPM fluid,
it is still essentially far from a quantitative agreement with the computer simulation results~\cite{Hynnien-Panagiotopoulos}.
This general disadvantage of the RPA  can be corrected by taking into account the effects of higher-order fluctuations.

In order to go beyond the RPA we use the method proposed
 for an ionic model in the bulk \cite{patsahan_ion,patsahan-mryglod-patsahan:06}. We start with the expression for $\beta\overline{P}_{\rm{G}}$
\begin{eqnarray*}
 \beta\overline{P}_{\rm{G}}=\beta\overline{P}^{r}-\frac{1}{2V}\sum_{{\mathbf{k}}}\ln\left[
 1+\rho_{1}\beta\widetilde{\varphi}^{C}(k)\right],
%\label{P_G}
\end{eqnarray*}
obtained with the use of (\ref{2.2}) and (\ref{Omega_rep}) under conditions
$\widetilde{\varphi}^{00}(k)=0$ and $\widetilde{\varphi}^{0I}(k)=0$. Here,  $\overline{P}^{r}=({\rm d}P^{r}/{\rm d}s)|_{s=0}$,
and $\rho_{1}=\bar\rho_{1}|_{s=0}$.
Both $\overline{P}^{r}$ and $\rho_{1}$ depend on the chemical potential $\overline{\mu}_{1}$. Presenting the chemical potential
as $\beta\overline{\mu}_{1}=\beta\overline{\mu}_{1}^{r}+\Delta\nu_{1}$, we expand $\beta\overline{P}_{\rm{G}}$
in powers of $\Delta\nu_{1}$ up
to the fourth power
\begin{equation*}
 \beta\overline{P}_{\rm{G}}=\sum_{n=0}^{4}\frac{{\cal M}_{n}}{n!}\Delta\nu_{1}^{n},
\end{equation*}
where ${\cal M}_{n}=\partial^{n}(\beta\overline{P}_{\rm{G}})/\partial\Delta\nu_{1}^{n}\vert_{\Delta\nu_{1}=0}$.
Following the procedure described in \cite{patsahan_ion,patsahan-mryglod-patsahan:06} for the bulk RPM, we get the  equation for
the vapour-liquid spinodal of the RPM confined in a random hard-sphere matrix
\begin{equation}
 \frac{1}{V}\sum_{{\mathbf{k}}}\widetilde{g}(k)=-2\frac{\overline{{\mathfrak{M}}}_{2}}{\overline{{\mathfrak{M}}}_{3}}
 +\frac{\overline{{\mathfrak{M}}}_{3}}{\overline{{\mathfrak{M}}}_{4}},
 \label{spinodal}
\end{equation}
where $\widetilde{g}(k)$ is given in (\ref{g_k}) and
\begin{equation}
S_{n}=\frac{\overline{{\mathfrak{M}}}_{n}}{\rho_{1}}=\frac{\partial^{n}\beta\overline{P}^{r}}{\partial(\beta\overline{\mu}_{1}^{r})^{n}},
 \qquad n\geq 2
\label{S_n}
\end{equation}
are  determined  from the thermodynamics of  the  reference system. In particular, $S_{2}$ is connected with
the isothermal compressibility of the reference system
\begin{equation}
 S_{2}=\rho_{1}k_{B}T\kappa_{T}=\Big[\Big(\frac{\partial\beta\overline{P}^{r}}
 {\partial\rho_{1}}\Big)_{T}\Big]^{-1}.
 \label{kappa}
\end{equation}
Explicit expressions for $S_{2}$, $S_{3}$ and $S_{4}$ are given in
Appendix.

\subsection{Thermodynamics of the reference system: The scaled-particle theory }

In order to describe the thermodynamic properties of the reference system needed in this perturbation scheme we use the SPT theory
reformulated recently for a hard-sphere fluid in a quenched hard-sphere matrix forming a random porous medium.  Here, we  only  outline the main ideas
of the theory  and present  the expressions for the chemical potential  and pressure of a hard-sphere fluid in a disordered matrix which
will be used in the current study.
One can find a detailed  presentation of the SPT theory in \cite{Holovko_Patsahan_Dong:12,Holovko_Patsahan_Dong:13}.

The basic idea of the SPT approach is an insertion of an additional scaled particle of a variable size into a fluid. The procedure of
the insertion   is equivalent to the creation of a cavity which is free of any other fluid particles. The key point of the
considered reformulation of the SPT theory consists in a derivation of the excess chemical potential of a scaled particle $\mu_{s}^{ex}$,
which is equal to work needed to introduce the particle of radius $r_s$ into a fluid in the presence of matrix particles.

An exact expression for a point scaled particle in a hard-sphere fluid in a random porous medium was obtained in~\cite{Hol09}. 
Combining this expression with the thermodynamic consideration of the large scaled particle,  
the chemical potential for a hard-sphere fluid in a hard-sphere matrix was obtained.
However, the approach proposed in~\cite{Hol09} referred to as SPT1 contains a subtle inconsistency appearing when the size of matrix particles is essentially
larger than the size of fluid particles. Later, this inconsistency was eliminated in a new approach named as SPT2~\cite{Pat11}.
Starting from this formalism a series of new approximations were developed~\cite{Pat11,Holovko_Patsahan_Dong:12,Holovko_Patsahan_Dong:13}.
Among these approximations we distinguish the SPT2b1 approximation which was approved as the most accurate one being in a very
good agreement with the Monte-Carlo simulation results~\cite{Kalyuzhnyi_Holovko_Patsahan_Cummings:14}.
The expressions obtained in SPT2b1 are used in our paper for the description of
the reference system. Consequently, we can write the following expression for the chemical potential $\overline{\mu}_{1}^{r}$
\begin{eqnarray}
(\beta \overline{\mu}_{1}^{r})^{SPT2b1}=\ln(\Lambda_1^3 \rho_1)-\ln(\phi)-\ln(1-\eta_{1}/\phi_{0})
+(1+A)\frac{\eta_{1}/\phi_{0}}{1-\eta_{1}/\phi_{0}}\nonumber\\
+\frac{\eta_{1}(\phi_{0}-\phi)}{\phi_{0}\phi(1-\eta_{1}/\phi_{0})}
+\frac12(A+2B)\frac{(\eta_{1}/\phi_{0})^{2}}{(1-\eta_{1}/\phi_{0})^{2}}+\frac{2}{3}B\frac{(\eta_{1}/\phi_{0})^{3}}
{(1-\eta_{1}/\phi_{0})^{3}}.
\label{hol2.18}
\end{eqnarray}
Here,  the following notations are introduced.  $\eta_{0}=\frac{\pi}{6}\rho_{0}\sigma_{0}^{3}$ and
$\eta_{1}=\frac{\pi}{6}\rho_{1}\sigma_{1}^{3}$ are the packing fractions of the matrix and fluid particles, respectively.
$\phi_{0}$ is the
geometrical porosity: $\phi_{0}=1-\eta_{0}$ . The second type of porosity $\phi$,
called the probe-particle porosity, is  defined by the excess
chemical potential of a fluid in the limit of infinite dilution \cite{Pat11}. It has the form:
\begin{eqnarray}
 \phi=(1-\eta_{0})\exp\left[-\left(\frac{3\eta_{0}\tau}{1-\eta_{0}}+
\frac{3\eta_{0}\left(1+\frac12\eta_{0}\right)\tau^{2}}{(1-\eta_{0})^{2}}+
\frac{\beta P_{0}\eta_{0}\tau^{3}}{\rho_{0}}\right)\right],
\label{hol2.6}
\end{eqnarray}
where $\tau=\sigma_{1}/\sigma_{0}$, and
\begin{equation*}
\frac{\beta P_{0}}{\rho_{0}}=\frac{(1+\eta_{0}+\eta_{0}^{2})}{(1-\eta_{0})^{3}}.
\end{equation*}
Coefficients $A$ and $B$, which determine  the porous medium structure, are as follows:
\begin{eqnarray*}
A&=&6+\frac{3\eta_{0}\tau(\tau+4)}{1-\eta_{0}}+\frac{9\eta_{0}^{2}\tau^{2}}{(1-\eta_{0})^{2}}\nonumber\\
B&=&\frac92\left(1+\frac{\tau\eta_{0}}{1-\eta_{0}}\right)^{2}
\end{eqnarray*}
The expression for  the pressure  obtained by using (\ref{hol2.18}) and the Gibbs-Duhem equation reads
\begin{eqnarray}
\left(\frac{\beta\overline{P}^{r}}{\rho_{1}}\right)^{SPT2b1}&=&-\frac{\phi_{0}}{\phi}\frac{1}{(1-\eta_{1}/\phi_{0})}+
\frac{\phi_{0}}{\eta_{1}}\left(\frac{\phi_{0}}{\phi}-1\right)
\ln\left(1-\frac{\eta_{1}}{\phi_{0}}\right)\nonumber\\
&&+\frac{A}{2}\frac{\eta_{1}/\phi_{0}}{(1-\eta_{1}/\phi_{0})^{2}}+\frac{2B}{3}\frac{(\eta_{1}/\phi_{0})^{2}}{(1-\eta_{1}/\phi_{0})^{3}}.
\label{hol2.19}
\end{eqnarray}

\section{Results and Discussions}

Using the theory described above we study the vapour-liquid phase diagram of the  RPM confined in a hard-sphere matrix depending
on the matrix properties (porosities $\phi_{0}$ and $\phi$) as well as on the parameter of size asymmetry
$\tau$. The calculations are performed in two approximations: the RPA and the approximation when the
correlation effects
of  higher  than the second order are taken into account (equation (\ref{spinodal})). We use the Weeks-Chandler-Andersen 
regularization scheme \cite{wcha}
for the Coulomb potential
inside the hard core. In this case,  $\widetilde\varphi^{C}(k)=4\pi q^{2}\sin(k\sigma_{1})/(\epsilon k^{3})$.  Hereafter,   the following
reduced units are introduced for the temperature $T^{*}=k_{B}T\epsilon\sigma_{1}/q^{2}$ and for the density $\rho^{*}=\rho\sigma_{1}^{3}$.

\subsection{Vapour-liquid binodals obtained in the RPA}

\begin{figure}[htb]
\begin{center}
\includegraphics [height=0.4\textwidth]  {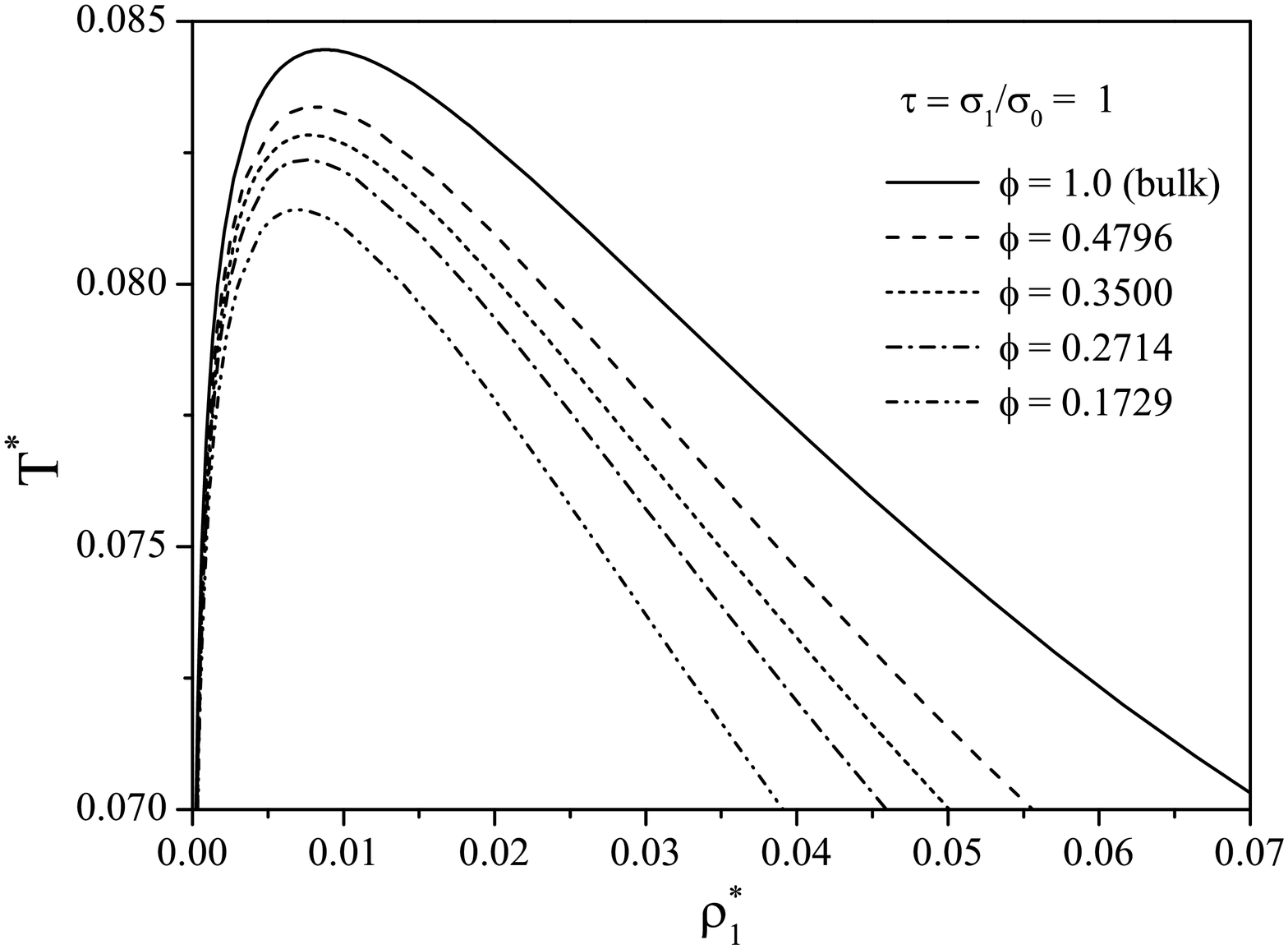}
\caption{Vapour-liquid coexistence curves for the RPM fluid confined in a hard-sphere matrix at different porosities,
but at the fixed size ratio $\tau=1$. The results are obtained in the RPA.}
\label{fig:Fig1}
\end{center}
\end{figure}

In order to calculate the coexistence curves we use equations~(\ref{nu_rpa})-(\ref{g_k}) and (\ref{hol2.18}) for
the chemical potential
and employ the Maxwell double-tangent construction.
First, we consider the binodals of vapour-liquid diagrams for the RPM fluid in the matrices of different probe-particle porosities,
$\phi$, but at the fixed size ratio of the fluid and matrix particles, $\tau=1$ (figure~\ref{fig:Fig1}).
As one can observe, both the critical temperature $T_c^{*}$ and the critical densities $\rho_c^{*}$ are lowering with the decrease of
the matrix porosity.
Simultaneously, the region of vapour-liquid coexistence is getting narrower.
This common behaviour is observed for many other types of fluids. Since our model does not contain any attractive interactions
between the fluid and matrix particles, only two most essential effects can play the role: the effect of excluded volume occupied by
the matrix particles and the weakening of attraction between the fluid particles due to the presence of matrix particles.
In comparison with the bulk fluid ($\phi=1.0$), the first effect leads to an increase of the fluid local density
that is why the total density can be smaller than in the bulk case. The second effect lowers an average coordination number of fluid 
particles
due to a contact of the fluid with a pore surface formed by the matrix particles, hence the effective attraction in such a system
is less than in the bulk. Thus, the critical temperature can decrease. This is typical of most liquids  and our results show that 
ionic liquids are not an exception,
at least for the RPM  within the RPA.
%%%%%%%%%%%%%%%%%%%%%%%%%%%%%%%%%%%%%%%%%%%%%%%%%%%%%%%%%%%%%%%%%%%%%%%%%%%%%%%%%

As it  was mentioned above, we distinguish two types of porosity: the probe-particle porosity $\phi$ and the geometrical porosity $\phi_0$.
A principal difference between them consists in the fact that the porosity $\phi$ takes into account the size of adsorbate particles,
while the porosity $\phi_0$ is an adsorbate-independent characteristic. Both of these porosities are important.
The geometrical porosity defines a ''bare'' pore volume of the matrix and it can be considered as a more general characteristic.
On the other hand, it does not take into account that some pores are inaccessible for the given fluid molecules due to their sizes:
a distance between the opposite pore walls can be less than the diameter of fluid particles.
By contrast, the probe-particle porosity $\phi$ defines a ''real'' pore volume, which is accessible for the fluid under study. 
%%%%%%%%%%%%%%%%%%%%%%%%%%%%%%%%%%%%%%%%%%%%%%%%%%%%%%%%%%%%%
\begin{figure}[htb]
\begin{center}
\includegraphics [height=0.35\textwidth] {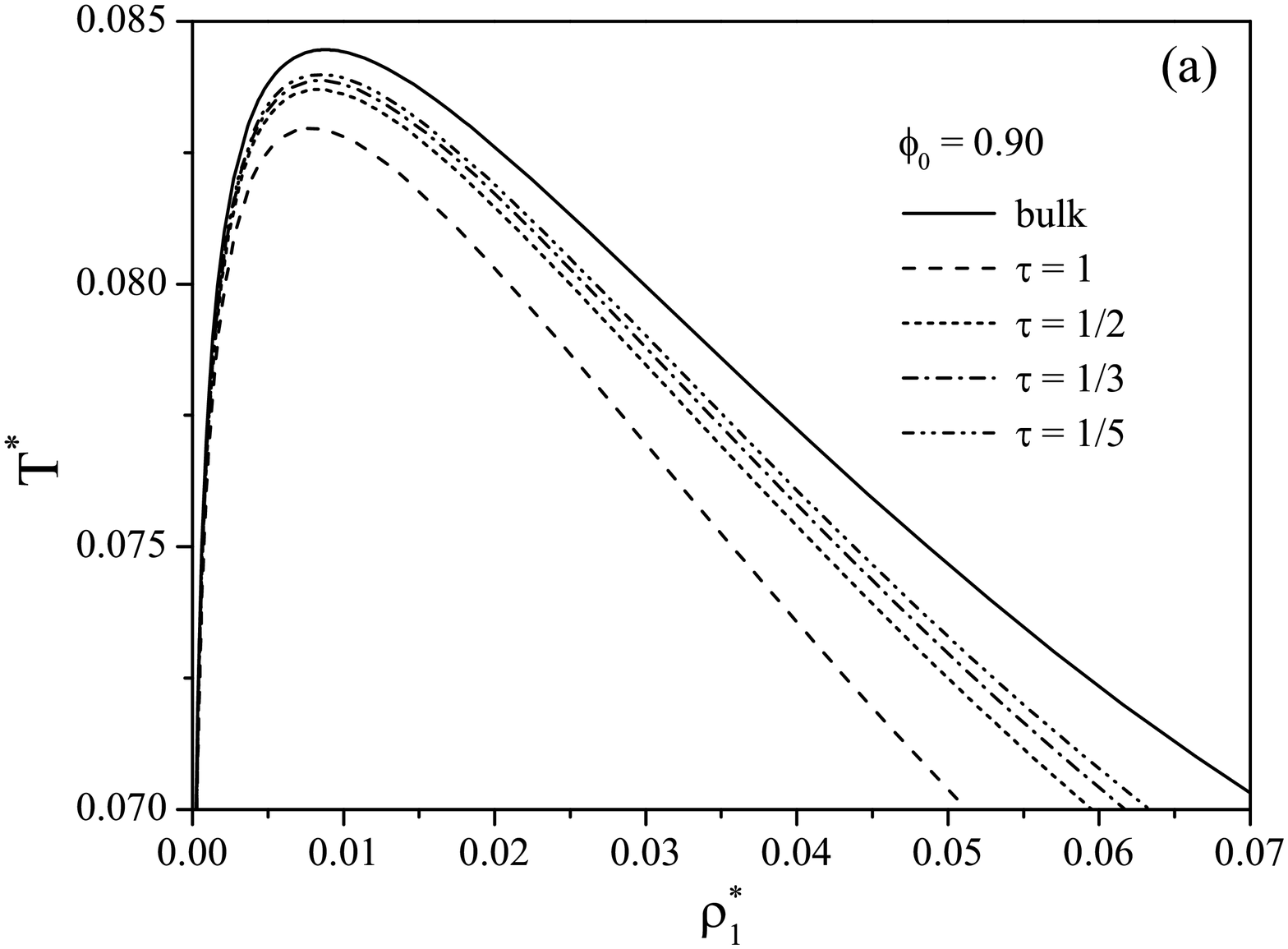}
\includegraphics [height=0.35\textwidth] {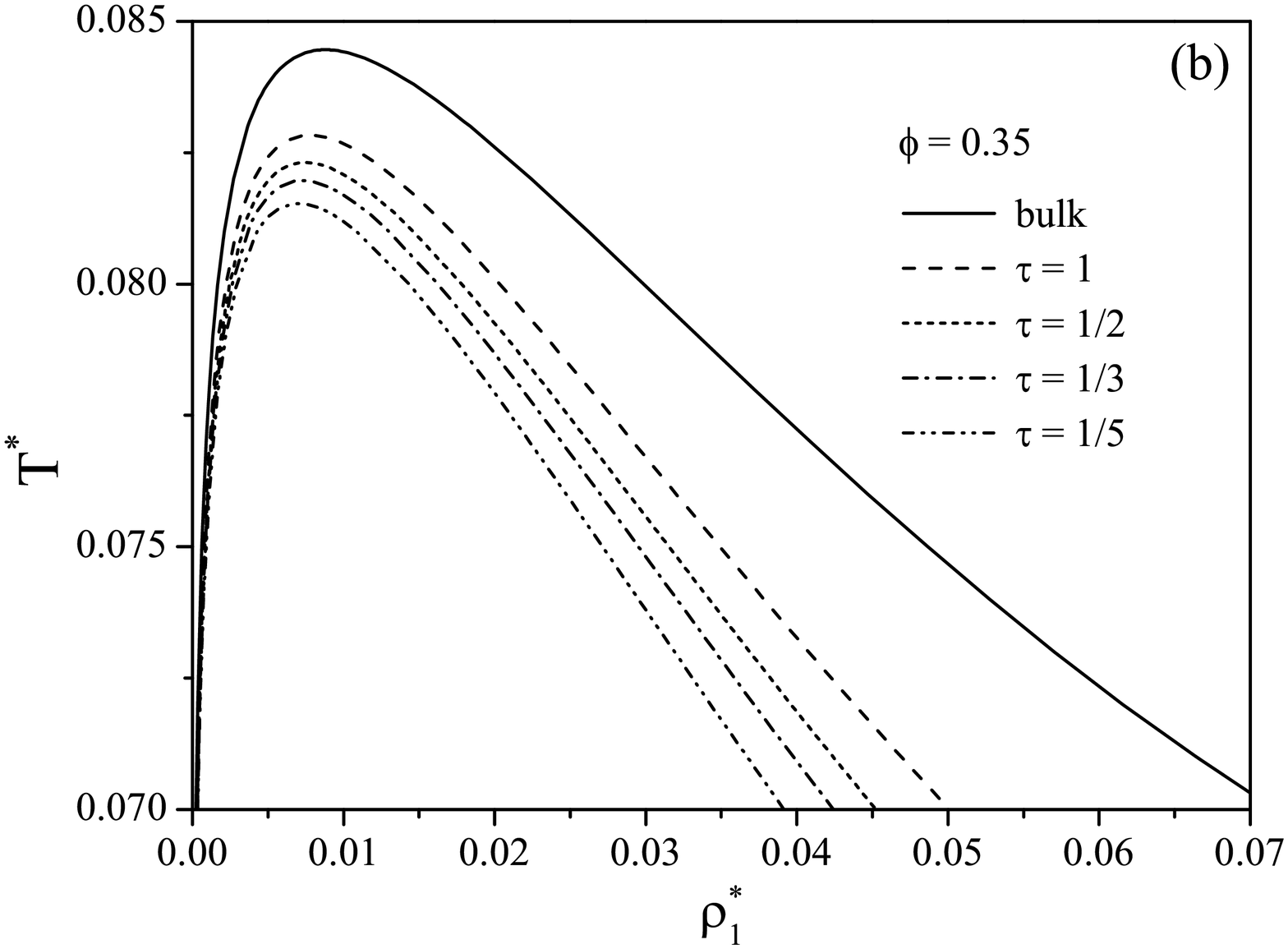}
\caption{Vapour-liquid coexistence curves for the RPM fluid confined in a hard-sphere matrix at different size ratios $\tau$, 
but at the fixed
porosities   $\phi_{0}=0.90$~(a) and $\phi=0.35$~(b). The results are obtained in the RPA.}
\label{fig:Fig2}
\end{center}
\end{figure}

Regardless of which type of the porosities is under control, a decrease both of them  leads to a decrease of the critical parameters
$T_{c}^{*}$ and $\rho_{c}^{*}$. However, the effect of the size ratio parameter $\tau$ is more ambiguous.
In figure~\ref{fig:Fig2} we present  vapour-liquid phase diagrams for the RPM fluid in matrices formed by
particles of different sizes, but at constant porosities $\phi_{0}$ (panel~a) and $\phi$ (panel~b).
As one can see, there are  opposite dependencies of critical parameters on the  ratio of fluid and matrix particles $\tau$.
To understand this behaviour it should be noted that the probe-particle porosity is always less than or equal to
the geometrical porosity ($\phi\leq\phi_0$). Moreover,  both porosities become equivalent in the limit $\tau\rightarrow0$,
when a bulk-like fluid at the effective density $\eta^{*}_1=\eta_1/\phi$ is obtained~\cite{Pat11}.
The bulk-like fluid can be reached due to a negligibly small effect of a pore surface in this limit: a specific pore surface area  
vanishes with respect to the specific free volume. Hence, the  fraction of inaccessible pores in the system tends to zero as well.
Therefore, if the porosity $\phi_0$ is fixed and the ratio $\tau$ decreases (figure~\ref{fig:Fig2}a),
then the porosity $\phi$ increases.
As it is shown in the previous figure, an increase of the porosity $\phi$ leads to an increase of $T_{c}^{*}$ and $\rho_{c}^{*}$.
In the case when the porosity $\phi$ is constant, but $\tau$ decreases (figure~\ref{fig:Fig2}b), one observes that
the critical temperature
and density decrease as well.
This seeming contradiction can be explained by the fact that  the specific
free volume which makes a greater contribution to the thermodynamics  of the system  than the specific pore surface area effect,  
decreases  in this case.
%%%%%%%%%%%%%%%%%%%%%%%%%%%%%%%%%%%%%%%%%%%%%%%%%%%%%%%%%%%%%%%%%%%%%%%%%%%%%%%%%%%%%%%%%%%%%%%%%%%%%%%%%%%%%%%%%%%%%%%%%%%%%%%%%%%%%%%%%%%%%%%%%%%%%
\begin{figure}[htb]
\begin{center}
\includegraphics [height=0.33\textwidth]  {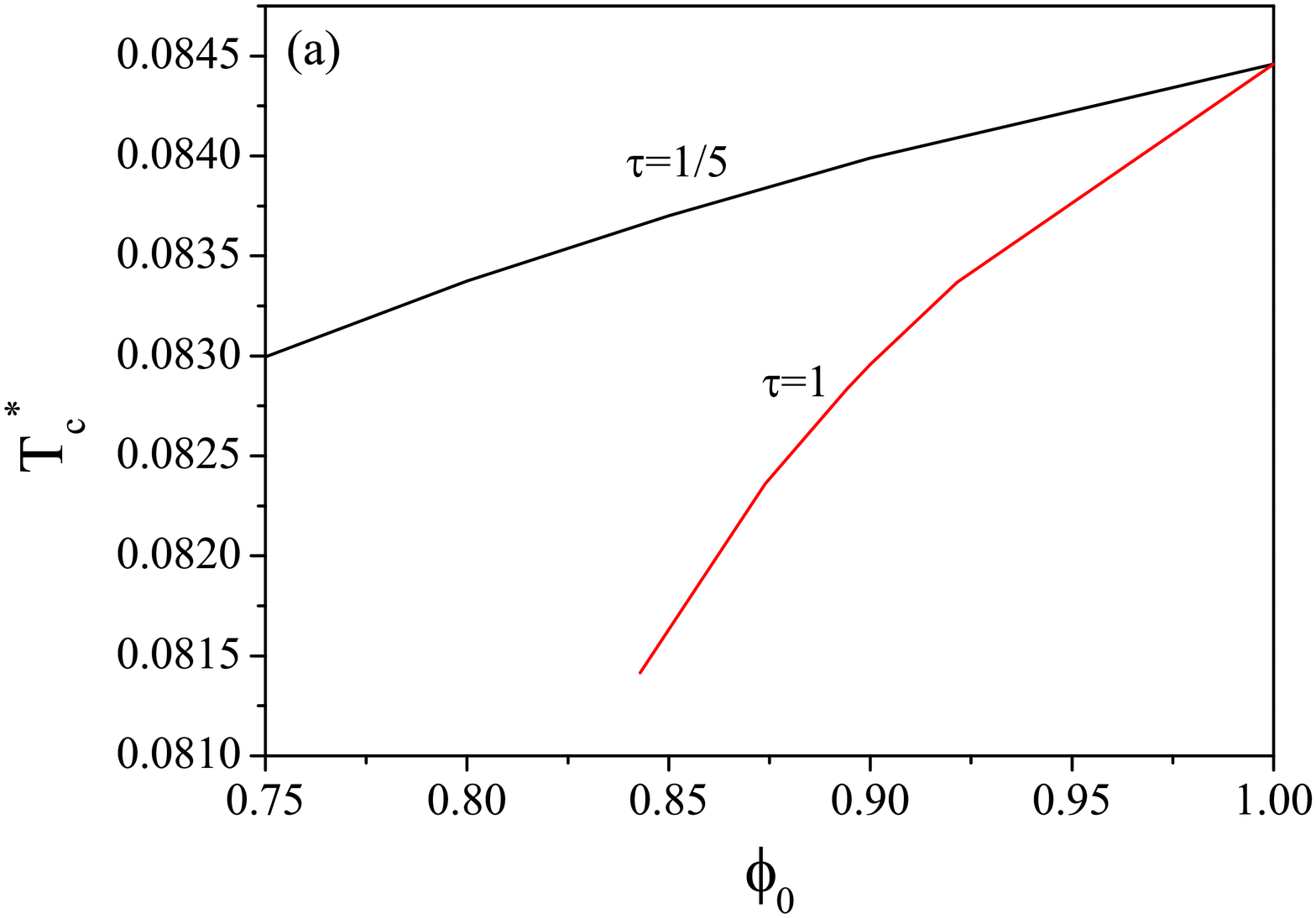}
\includegraphics [height=0.34\textwidth]  {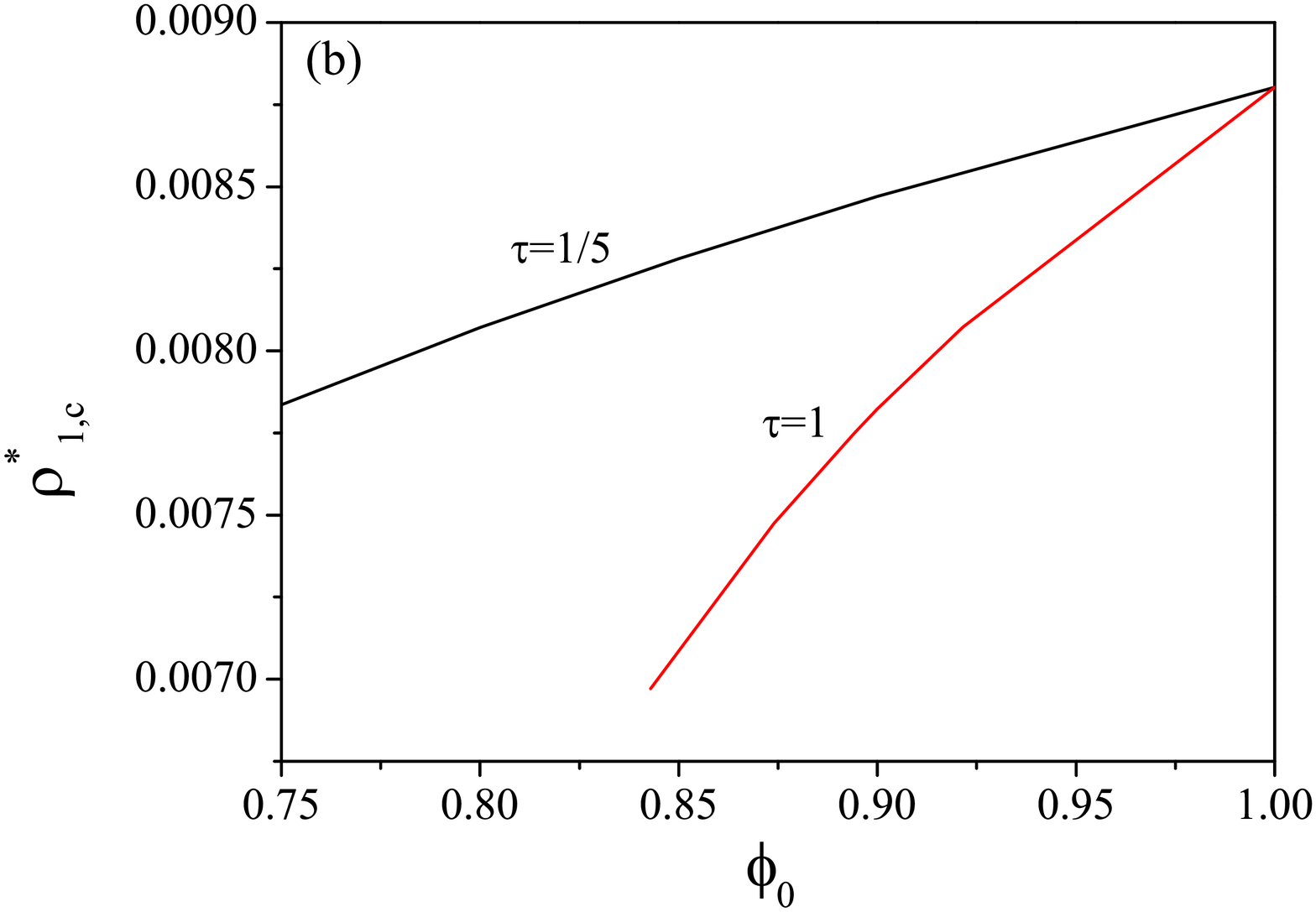}
\caption{Critical temperature $T_c^{*}$ (a) and critical density $\rho_{1,c}^{*}$ (b) of the RPM fluid confined in a hard-sphere matrix 
depending on the matrix porosity $\phi_0$. The results are obtained in the RPA.}
\label{fig:Fig3}
\end{center}
\end{figure}
%%%%%%%%%%%%%%%%%%%%%%%%%%%%%%%%%%%%%%%%%%%%%%%%%%%%%%%%%%%%%%%%%%%%%%%%%%%%%%%%%%%%%%%%%%%%%%%%%%%%%%%%%%%%

Based on the results presented in figure~\ref{fig:Fig2}a
we demonstrate  quantitative estimations of the critical parameters $T_{c}^{*}$ and $\rho_{c}^{*}$
depending on the porosity $\phi_0$ for the small and larger matrix particles (figure~\ref{fig:Fig3}).
It is clearly seen that the both critical parameters increase monotonously with increasing  $\phi_0$.
Naturally, for  large matrix particles ($\tau=1/5$) $T_{c}^{*}$ and $\rho_{c}^{*}$ are higher than in the case of
small matrix particles ($\tau=1$) everywhere until the bulk case is obtained ($\phi_0=1$), where
they  become equal.

\subsection{Vapour-liquid spinodals obtained with taking into account higher-order terms}

In our previous studies of a vapour-liquid phase transition of the RPM fluid in the bulk~\cite{patsahan_ion} we demonstrated
that the higher-order terms being taken into account essentially improved the results  and led to the value of the  
critical temperature $T_{c}^{*}=0.0502$, which is rather close to the one obtained from the computer simulations 
($T_{c}^{*}=0.049$) \cite{Hynnien-Panagiotopoulos}.
In comparison, the RPA  strongly overestimates the critical temperature by giving $T^{*}_{c}=0.084$ \cite{Cai-Mol1,patsahan_ion}.
The critical density obtained in the above-mentioned approximation is still underestimated ($\rho_{c}^{*}=0.044$) when compared with
the  simulation results ($\rho^{*}_{c}=0.06-0.08$) \cite{Hynnien-Panagiotopoulos}, but it is much better than in the case
of the RPA  giving $\rho^{*}_c\simeq 0.01$ \cite{Cai-Mol1,patsahan_ion}. Such an improvement achieved when the effects of higher-order
correlations are taken into account
has motivated us to extend this approach to the case of the RPM fluid confined in a hard-sphere matrix.
%%%%%%%%%%%%%%%%%%%%%%%%%%%%%%%%%%%%%%%%%%%%%%%%%%%%%%%%%%%%%%%%%%%%%%%%%%%%%%%%%%%%%%%%%%%%%%%%%%%%%%%%%%%%%%%%%%%%%%%%%%%%%%%%%%%%%%%%%%%%%%
\begin{figure}[htb]
\begin{center}
\includegraphics [height=0.4\textwidth]  {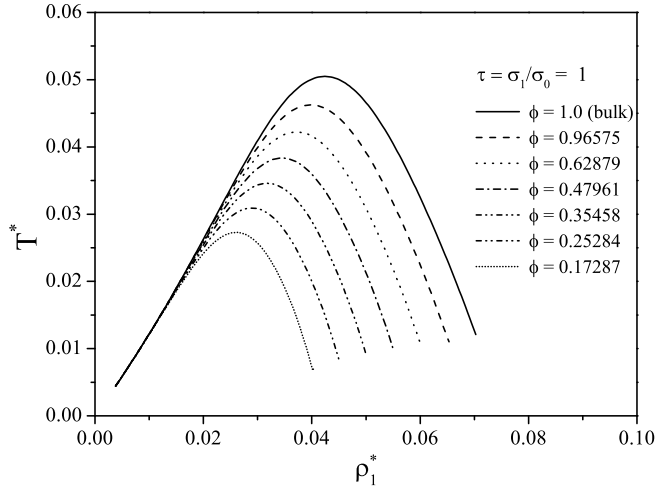}
\caption{Vapour-liquid spinodal curves for the RPM fluid confined in a hard-sphere matrix at different porosities,
but at the fixed size ratio $\tau=1$. The results are obtained by using equation (\ref{spinodal}).}
\label{fig:Fig4}
\end{center}
\end{figure}

Using (\ref{spinodal}) and (\ref{g_k}) in a combination with equations~(\ref{A_1})-(\ref{A_6}) we have calculated spinodals
for the vapour-liquid phase transition of the RPM fluid in matrices of  different porosities $\phi$, but
at the constant value of $\tau=1$~(figure~\ref{fig:Fig4}). From the obtained results one can observe the same qualitative
dependence of the critical temperature and density on the matrix porosity, i.e.  both parameters decrease
with a decrease of $\phi$. However, it is seen that in comparison with the RPA results (figure~\ref{fig:Fig1}) the critical
parameters obtained in the  approximation given by (\ref{spinodal}) are more sensible to the variation of  matrix porosity.
%%%%%%%%%%%%%%%%%%%%%%%%%%%%%%%%%%%%%%%%%%%%%%%%%%%%%%%%%%%%%%%%%%%%%%%%%%%%%%%%%%%%%%%%%%%%%%%%%%%%%%%%%%%%%%%%%%%%%%%%
\begin{figure}[htb]
\begin{center}
\includegraphics [height=0.35\textwidth]  {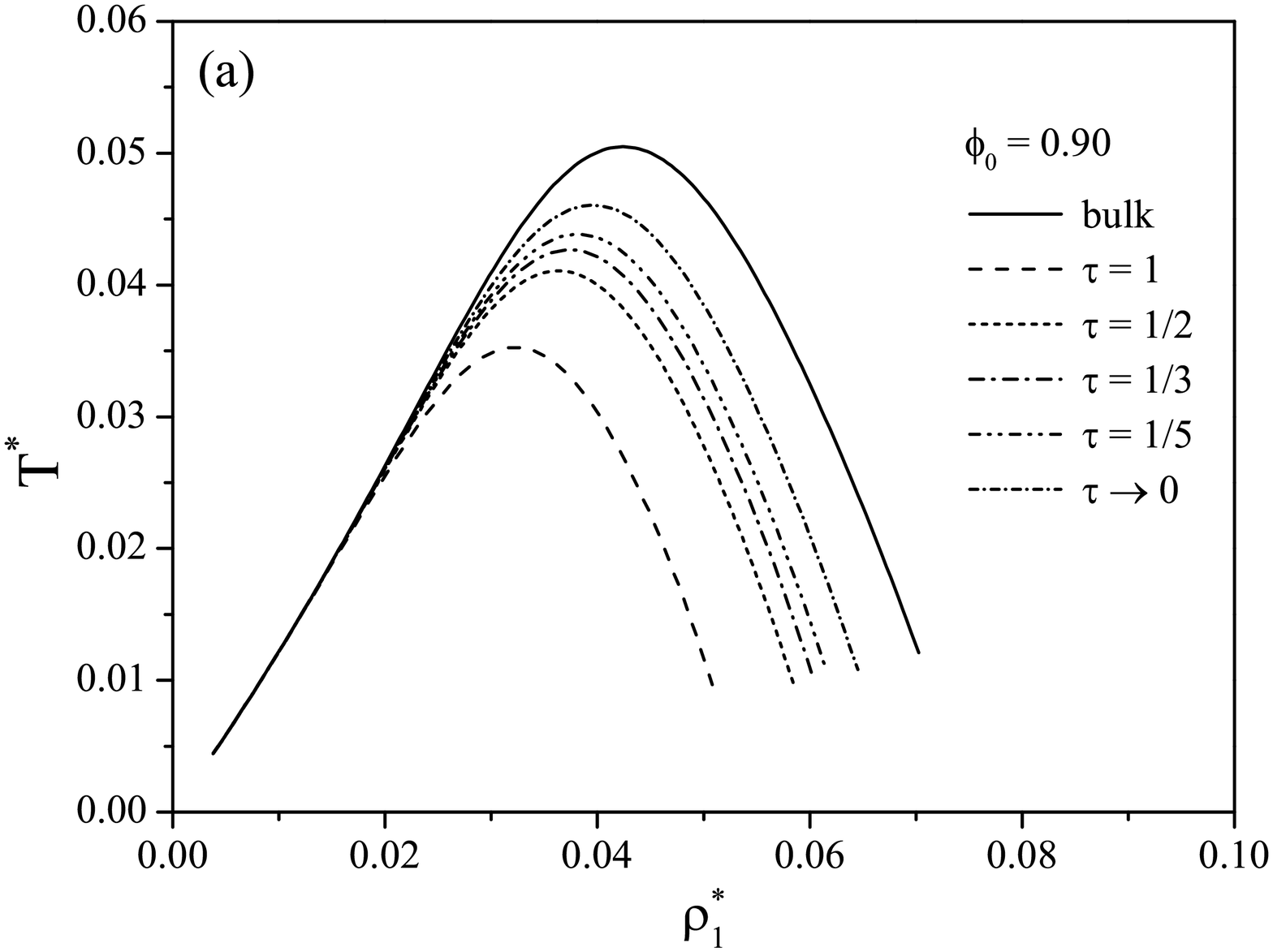}
\includegraphics [height=0.35\textwidth]  {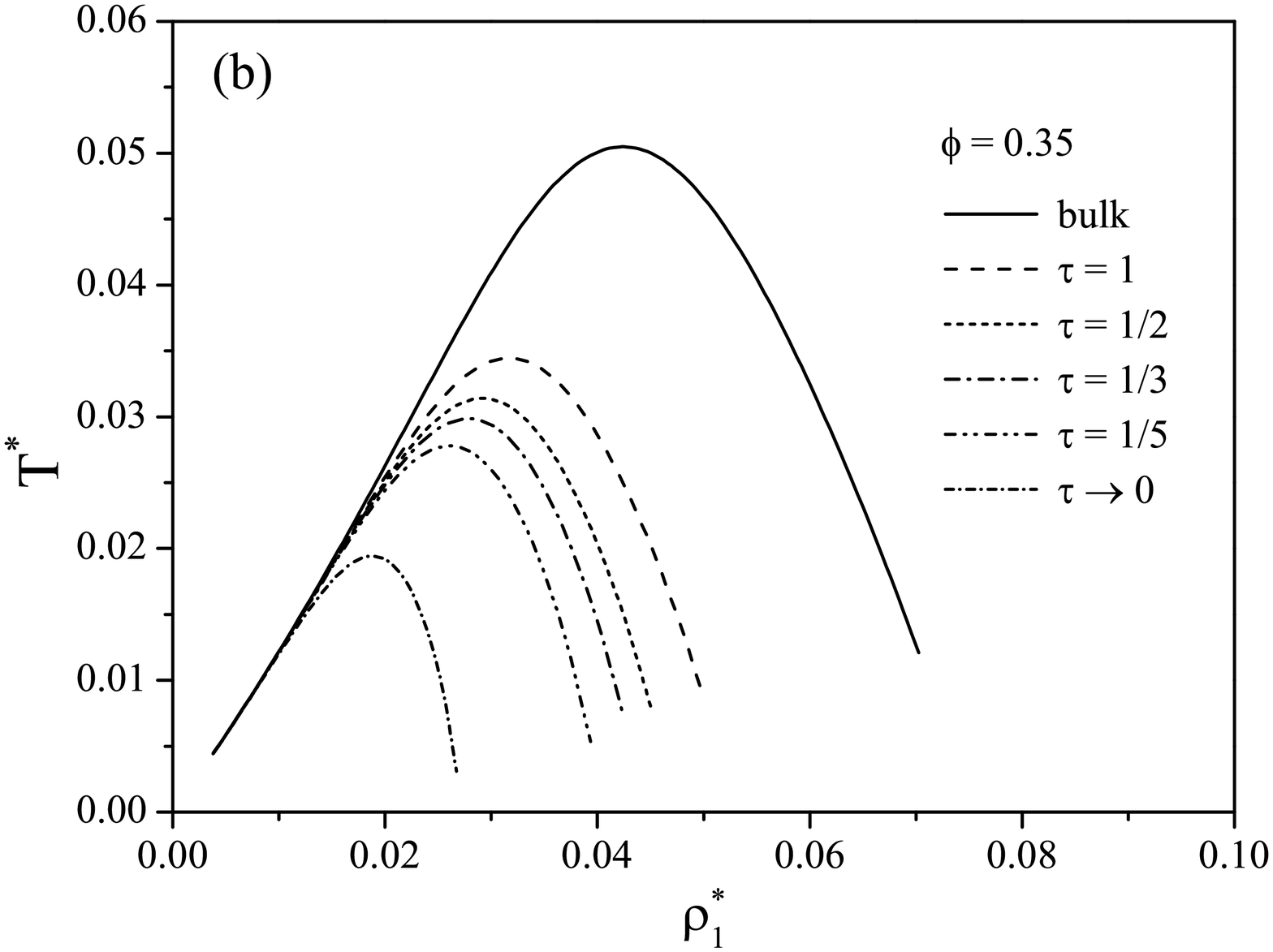}
\caption{Vapour-liquid spinodal curves for the RPM fluid confined in a hard-sphere matrix at different size ratios $\tau$, but at the  fixed
porosities $\phi_{0}=0.90$ (a) and $\phi=0.35$ (b). The results are obtained by using equation (\ref{spinodal}).}
\label{fig:Fig5}
\end{center}
\end{figure}

The effect  of the matrix particle size at fixed porosities $\phi_0=0.90$ and $\phi=0.35$ is shown in figure~\ref{fig:Fig5}.
For these cases, one can notice the same tendencies as the ones obtained in the RPA, but again  both
critical parameters are lowering faster with a decrease of $\tau$.
Therefore, it should be acknowledged that this sensibility to the matrix porosity is directly related to
taking into account the higher-order terms.
%%%%%%%%%%%%%%%%%%%%%%%%%%%%%%%%%%%%%%%%%%%%%%%%%%%%%%%%%%%%%%%%%%%%%%%%%%%%%%%%%%%%%%%%%%%%%%%%%%%%%%%%%%%%%%%%%%%%%%%%%%%%%
\begin{figure}[htb]
\begin{center}
\includegraphics [height=0.34\textwidth]  {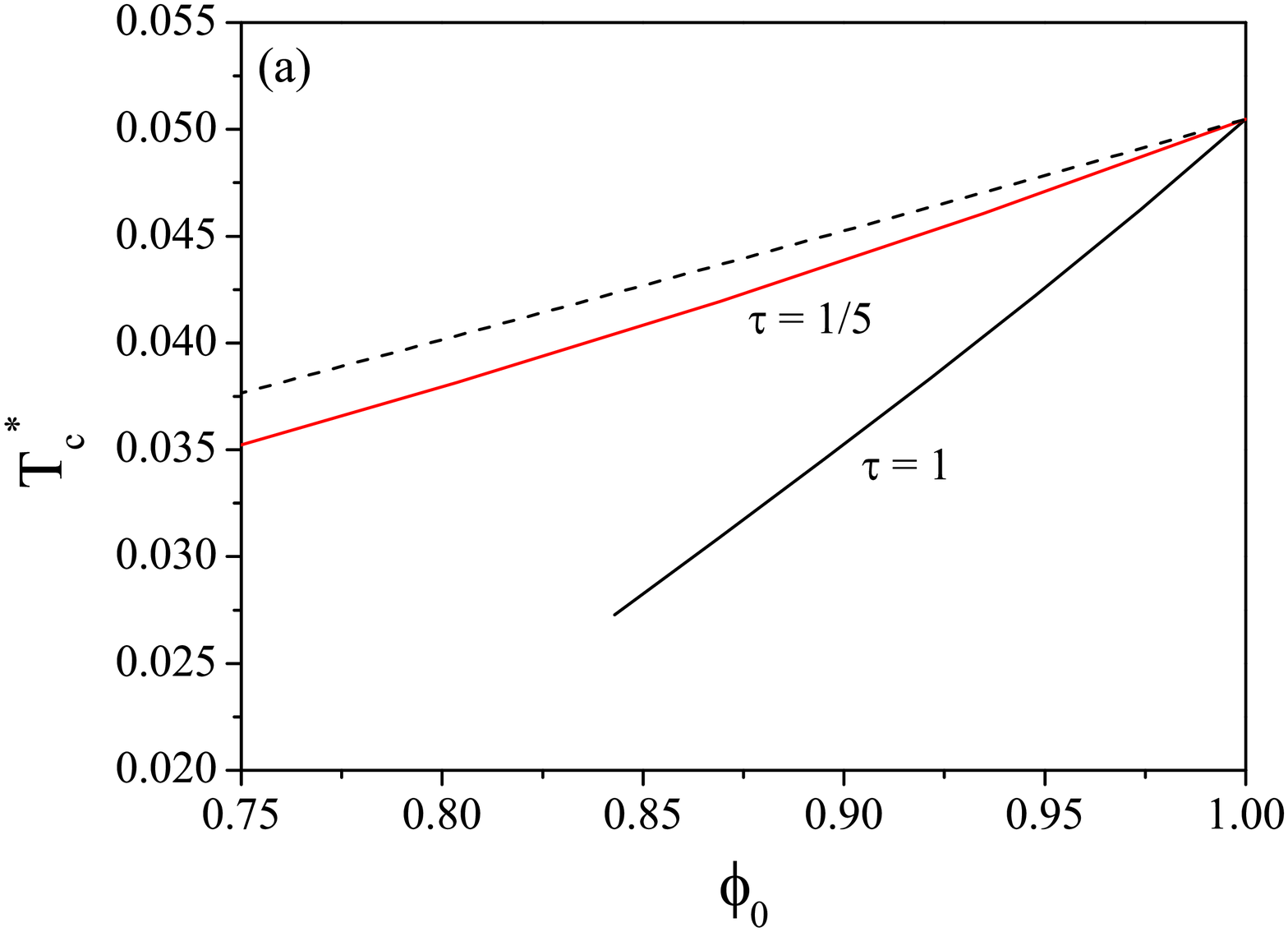}
\includegraphics [height=0.34\textwidth]  {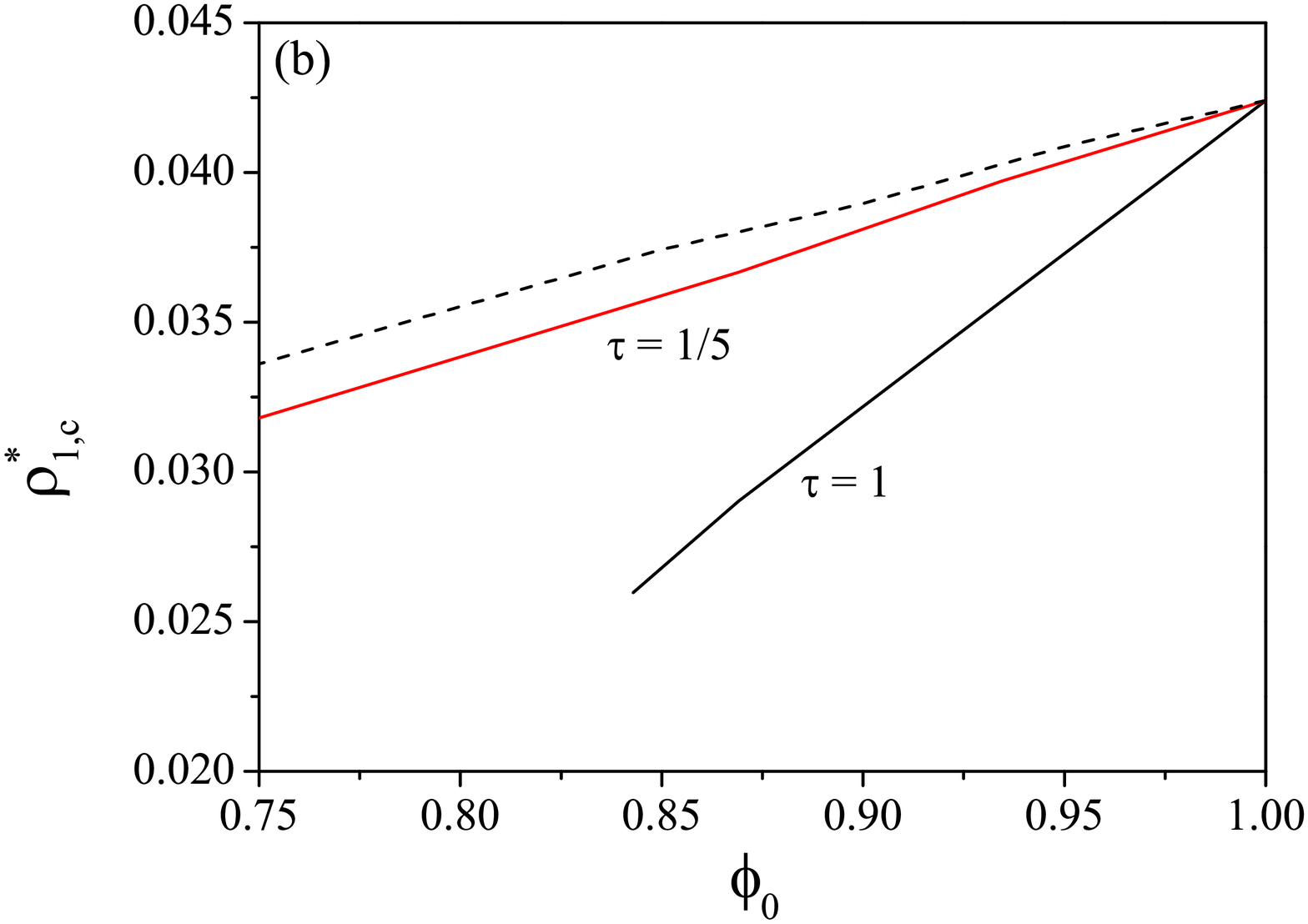}
\caption{Critical temperature $T_c^{*}$ (a) and critical density $\rho_{1,c}^{*}$ (b) of the RPM fluid confined in a 
hard-sphere matrix depending on the matrix porosity $\phi_0$. Dashed line corresponds to the 
limit $\tau\rightarrow0$ (or $\sigma_0/\sigma_1\rightarrow\infty$). The results  are obtained by using equation (\ref{spinodal}).}
\label{fig:Fig6}
\end{center}
\end{figure}

The quantitative analysis   of the critical temperature and density behaviour shows
not only their rapid and monotonous variation with respect to the matrix porosity $\phi_0$ (figure~\ref{fig:Fig6})
in comparison with the RPA results~(figure~\ref{fig:Fig3}), but also
these dependencies are very close to linear ones. This is one more distinction between the RPA
and the results obtained in the higher-order approximation.
As usual, $T_c^{*}$ and $\rho_c^{*}$ for smaller matrix particles ($\tau=1$) are lower than for larger ones ($\tau=1/5$).
Besides that, we have estimated the limit case $\tau\rightarrow0$ (infinite size of matrix particles) shown
by  dashed lines in figure~\ref{fig:Fig6} which is  the upper bound of the corresponding critical parameters
at the given matrix porosities.
%%%%%%%%%%%%%%%%%%%%%%%%%%%%%%%%%%%%%%%%%%%%%%%%%%%%%%%%%%%%%%%%%%%%%%%%%%%%%%%%%%%%%%%%%%%%%%%%%%%%%%%%%%%%%%%%
\begin{figure}[htb]
\begin{center}
\includegraphics [height=0.35\textwidth]  {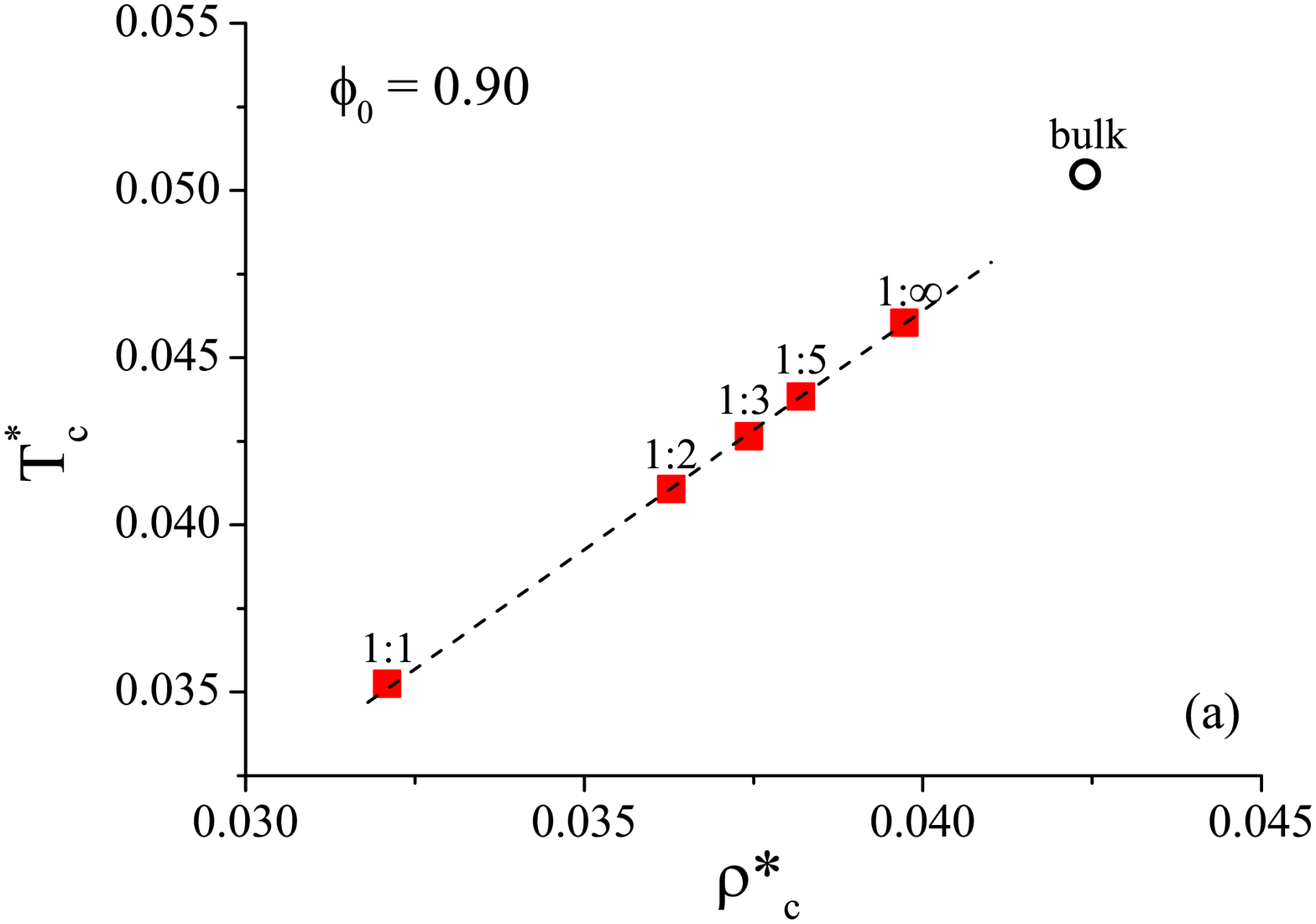}
\includegraphics [height=0.35\textwidth]  {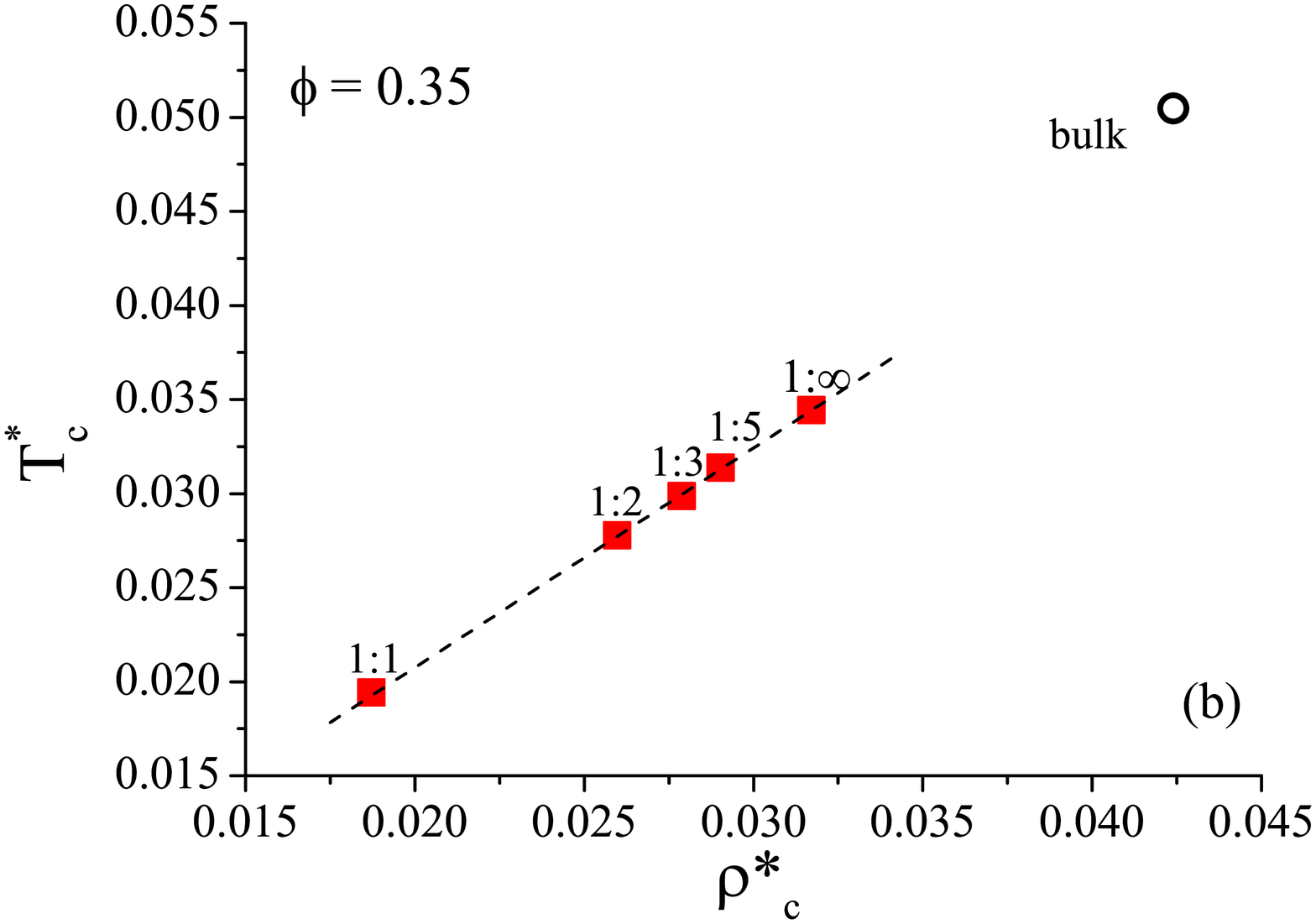}
\caption{Critical parameters of the RPM fluid confined in a hard-sphere matrix at different size ratios $\sigma_1$:$\sigma_0$ 
(solid symbols), but at the fixed
porosities $\phi_{0}=0.90$ (a) and $\phi=0.35$ (b). Open symbol corresponds to the RPM fluid in the bulk.
The results are obtained by using equation (\ref{spinodal}).}
\label{fig:Fig7}
\end{center}
\end{figure}
%%%%%%%%%%%%%%%%%%%%%%%%%%%%%%%%%%%%%%%%%%%%%%%%%%%%%%%%%%%%%%%%%%%%%%%%%%%%%%%%%%%%%%%%%%%%%%%%%

Finally, we have built plots for the critical parameters in the $T_c^{*}$-$\rho_c^{*}$ coordinates
for the porosities $\phi_0=0.90$ (figure~\ref{fig:Fig7}a) and $\phi=0.35$ (figure~\ref{fig:Fig7}b) and for different values of
ratio parameter $\tau$.
One can notice  interesting dependencies for the both cases, where a linear behaviour is obtained. The same type of
behaviour is found for
the critical parameters obtained in the RPA.

\section{Conclusions}
We have studied the effects of disordered porous media on the vapour-liquid phase behaviour of a model ionic fluid consisting of an
electroneutral equimolar mixture of  equisized charged hard spheres immersed in a structureless dielectric continuum
(the so-called RPM model).
To this end, based on an extended  Madden-Gland model,  we have developed a theoretical approach which combines the method of CVs
with  the SPT theory
reformulated recently for a hard-sphere fluid  confined  in a disordered hard-sphere matrix. While the CV theory appears to be
successful  in  studying the phase behaviour of the systems with Coulomb interactions, the SPT provides an accurate analytical
description of the thermodynamic properties of the reference system, i.e., the system including  a hard-sphere repulsion.
We have formulated a theory for the system
in which,  in addition
to a hard-core repulsion,  
pair  (repulsive/attractive) interactions between the matrix particles and between the ions and the matrix particles are
taken into account.

In this contribution,   we  have obtained
an analytical expression for    the chemical potential of a model ionic fluid in a disordered matrix formed by uncharged hard spheres
in the RPA. Based on this expression, we have calculated the vapor-liquid binodals
and have analyzed the effects of the   matrix characteristics (geometrical and probe-particle porosities and a size ratio
of  matrix and fluid particles) on the coexistence
envelope. Also, using
the method proposed previously for the bulk RPM, we have derived an explicit equation for the vapour-liquid spinodal
which  takes into account the effects of the third- and fourth-order correlations.  Since this method allows us to obtain a good
quantitative agreement with simulation data for the critical parameters of an unconfined RPM, we suppose that it  has produced reasonable
quantitative  results for the system under consideration.
Both approximations have yielded the same  dependence  of  the vapour-liquid phase diagram on the matrix characteristics,
i.e., with a decrease of porosity the critical point shifts toward lower fluid densities and lower temperatures and the region of 
coexistence is getting narrower.
It has also been  observed that for the fixed matrix porosity, both the critical temperature and the critical density
increase with an increase of the size of matrix particles and tend to the critical  values found for the bulk RPM.

In conclusion,  the vapour-liquid behaviour of the RPM confined in  random hard-sphere matrices characterized by different porosities
and by different hard-sphere diameters has been studied for the first time. The RPM is the simplest model of ionic fluids which does
not take into account the charge, size and shape asymmetry of ions being usually typical of real ionic systems. Thus, the theory should be
developed to be capable of describing  phase diagrams of more complex models.  The work
in this direction is  in progress.

\appendix
\section*{Appendix}
\setcounter{section}{1}
\setcounter{section}{1}
Quantities $S_{3}$ and $S_{4}$ given by (\ref{S_n}) can be presented in terms of the derivatives with respect  to $\eta_{1}$
($\eta_{1}=\frac{\pi}{6}\rho_{1}\sigma_{1}^{3}$) as follows:
\begin{eqnarray}
S_{3}&=&S_{2}\left(S_{2}+\eta_{1}\frac{\partial S_{2}}
{\partial\eta_{1}}\right),
\label{A_1} \\
S_{4}&=&S_{2}\left[S_{2}^{2}+4S_{2}
\eta_{1}\frac{\partial S_{2}}{\partial\eta_{1}}+\eta_{1}^{2}\left(\frac{\partial S_{2}}
{\partial\eta_{1}}\right)^{2}
+S_{2}\eta_{1}^{2}\frac{\partial^{2}S_{2}}{\partial\eta_{1}^{2}}
 \right].
\label{A_2}
\end{eqnarray}
Then, one can find explicit expressions for $S_{2}$, $S_{3}$ and $S_{4}$ by using   the formulae
\begin{eqnarray}
S_{2}&=&\left[\frac{1}{1-\eta_{1}/\phi_{0}}+\frac{\eta_{1}}{\phi(1-\eta_{1}/\phi_{0})^{2}}
\right.
\nonumber
\\
&&
\left.
+\frac{A\eta_{1}}
{\phi_{0}(1-\eta_{1}/\phi_{0})^{3}}+\frac{2B\eta_{1}^{2}}{\phi_{0}^{2}(1-\eta_{1}/\phi_{0})^{4}} \right]^{-1},
\label{A_3}
\end{eqnarray}
\begin{equation}
\frac{\partial S_{2}}{\partial\eta_{1}}=-GS_{2}^{2}, \qquad
\frac{\partial^{2}S_{2}}{\partial\eta_{1}^{2}}=-FS_{2}^{2}+2G^{2}S_{2}^{3},
\label{A_4}
\end{equation}
\begin{eqnarray}
G=\frac{1}{\phi_{0}(1-\eta_{1}/\phi_{0})^{2}}+\frac{1+\eta_{1}/\phi_{0}}{\phi(1-\eta_{1}/\phi_{0})^{3}}+\frac{A(1+2\eta_{1}/\phi_{0})}
{\phi_{0}(1-\eta_{1}/\phi_{0})^{4}}\\ \nonumber
+\frac{4B\eta_{1}(1+\eta_{1}/\phi_{0})}{\phi_{0}^{2}(1-\eta_{1}/\phi_{0})^{5}},
\label{A_5}
\end{eqnarray}
\begin{eqnarray}
F=\frac{2}{\phi_{0}^{2}(1-\eta_{1}/\phi_{0})^{3}}+\frac{2(2+\eta_{1}/\phi_{0})}{\phi\phi_{0}(1-\eta_{1}/\phi_{0})^{4}}
+\frac{6A(1+\eta_{1}/\phi_{0})}
{\phi_{0}^{2}(1-\eta_{1}/\phi_{0})^{5}} \nonumber \\
+\frac{4B(1+\eta_{1}/\phi_{0})}{\phi_{0}^{2}(1-\eta_{1}/\phi_{0})^{5}}+\frac{8B\eta_{1}(3+2\eta_{1}/\phi_{0})}
{\phi_{0}^{3}(1-\eta_{1}/\phi_{0})^{6}}
\label{A_6}
\end{eqnarray}
which are obtained from (\ref{kappa}) and (\ref{hol2.19}).

%\newpage
%\section*{References}

\end{document}